\documentclass[epj,final]{svjour}
\usepackage{graphicx,setspace}
\usepackage{amsfonts,amsmath,amssymb,bm}
\usepackage[dvips,usenames]{color}
\begin{document}
\title{Phoretic Motion of Spheroidal Particles Due To 
Self-Generated Solute Gradients}
%\titlerunning{Spheroidal diffusiophoretic self-propellers}

\author{M. N. Popescu\inst{1} \and S. Dietrich\inst{2,3} \and 
M. Tasinkevych\inst{2,3} \and J. Ralston\inst{1}
}                     % Do not remove
%\authorrunning{Popescu, Dietrich, Tasinkevych, and Ralston}%
\mail{M. N. Popescu}
\institute{Ian Wark Research Institute, University of South Australia,
5095 Adelaide, South Australia, Australia\\ 
\email{Mihail.Popescu@unisa.edu.au}\\
\email{John.Ralston@unisa.edu.au}
\and
Max-Planck-Institut f\"ur Metallforschung, Heisenbergstr. 3,
70569 Stuttgart, Germany 
\and 
Institut f\"ur Theoretische und Angewandte Physik,
Universit\"at Stuttgart, Pfaffenwaldring 57, 70569 Stuttgart,
Germany\\
\email{dietrich@mf.mpg.de}\\
\email{miko@mf.mpg.de}
}
\date{Received: date / Revised version: date}
\abstract{
We study theoretically the phoretic motion of a spheroidal particle,
which generates solute gradients in the surrounding unbounded solvent
via chemical reactions active on its surface in a cap-like region
centered at one of the poles of the particle. We derive, within the 
constraints of the mapping to classical diffusio-phoresis, an 
analytical expression for the phoretic velocity of such an object. 
This allows us to analyze in detail the dependence of the velocity 
on the aspect ratio of the polar and the equatorial diameters of the 
particle and on the fraction of the particle surface contributing to 
the chemical reaction. The particular cases of a sphere and of an 
approximation for a needle-like particle, which are the most common 
shapes employed in experimental realizations of such self-propelled 
objects, are obtained from the general solution in the limits that 
the aspect ratio approaches one or becomes very large, respectively.
} %end of abstract
\maketitle
\section{Introduction}\label{intro}
The increasing interest in the development of ``lab on a chip'' 
devices and of drug-delivery systems has led to a stringent need of 
scaling standard machinery down to micro- and nano-scales. This 
reduction in length scale has raised a number of challenging issues, 
such as developing ways to enable small objects to perform 
autonomous, directional motion \cite{Whitesides_2002,Paxton_2004}.

Although the experimental and theoretical research in this area
is still in its early stages, several proposals for such
``self-propellers'' have already been tested experimentally (see,
e.g., Refs. 
\cite{Whitesides_2002,Paxton_2004,Sen_2005,Howse_2007,Leiderer_2008}); 
a review of the recent progress in this field
can be found in Ref. \cite{Paxton_review_2006}. These ``proof of
principle'' proposals have generally employed particles with axial
symmetry, i.e., cylindrical rods \cite{Whitesides_2002,Paxton_2004} 
or spheres \cite{Howse_2007,Leiderer_2008}, because they are 
relatively easy to manufacture, allow for a good control of the 
desired surface modifications, and their simple geometry is a 
significant bonus for the theoretical analysis of the experimental 
results. The underlying idea, as put forward by Whitesides and 
co-workers \cite{Whitesides_2002}, is that an asymmetric decoration 
of the particle with a catalyst, which promotes an activated 
reaction in the surrounding liquid medium generating product 
molecules, can provide motility through a variety of mechanisms. 
As in the initial design proposed in Ref. \cite{Whitesides_2002}, 
the simplest example is the propulsion of mm-size objects due to 
the ejection and subsequent bursting of bubbles formed by the 
product molecules [O$_2$ for PDMS plates with Pt catalyst tails 
placed in hydrogen peroxide (H$_2$O$_2$) aqueous solutions]. As the 
size of the particle is decreased towards the micron scale or below, 
viscous and surface forces start to dominate and inertia-based 
mechanisms such as the ``bubble ejection'' propulsion become 
ineffective. If the product molecules remain dissolved in the 
surrounding liquid medium as, e.g., in the experiments reported in 
Refs. \cite{Paxton_2004,Howse_2007,Leiderer_2008}, the result of an 
asymmetric distribution of catalyst is that the chemical reaction 
gives rise to concentration gradients along the surface of the 
particle. It has therefore been argued 
\cite{Paxton_2004,Golestanian_2005} that in such cases the motion of 
the catalyst-covered ``active'' particle is rather phoretic, i.e., 
the result of the interactions between the particle and the 
non-uniformly distributed product molecules generated by the chemical 
reaction. For example, in the case of Au-Pt rods in 
H$_2$O$_2$-H$_2$O mixtures \cite{Paxton_2004} the product molecules 
O$_2$ play the role of a solute the concentration gradient of which 
in the solution (formed by the H$_2$O$_2$ - H$_2$O mixture as the 
solvent and O$_2$ as the solute) is the field that may induce 
phoretic motion. 

In many cases the magnitude and the direction of the experimentally 
observed phoretic velocity of such ``active'' particles are compatible 
with a variety of microscopic mechanisms, such as surface
tension gradients \cite{Paxton_2004,Sen_2005,Saidulu_2008} (note that 
Ref. \cite{Sen_2005} provides an elegant example that rotational 
motion can also be achieved), cyclic adsorption and 
desorption \cite{Paxton_review_2006}, electrokinetics 
\cite{Paxton_2004,Leiderer_2008,Paxton_review_2006,Paxton_2005}, or
diffusio-phoresis 
\cite{Howse_2007,Golestanian_2005,Kapral_2007,Golestanian_2007}. 
(By using the notion of diffusio-phoresis, here we refer 
strictly to phoresis due to gradients of a solute, i.e., we do not 
consider the case of bimetallic particles 
\cite{Paxton_2004,Paxton_review_2006} for which charge transfer 
and electro-chemistry may be the dominant effects. For simplicity, 
we focus here on the case of electrically neutral solutes, which in 
the literature is often also called ``chemo-phoresis''.) 
Thus understanding these systems and discriminating between these
various possibilities require a careful, detailed theoretical
analysis to predict the dependence of the velocity on the control
parameters of the system, such as, e.g., the 
$\mathrm{H}_2\mathrm{O}_2$ content of the aqueous hydrogen peroxide
solvent used in the experiments reported in 
Refs. \cite{Paxton_2004,Howse_2007,Leiderer_2008} or the fraction 
of the particle surface which is catalytically active. With the 
notable exception of Ref. \cite{Kapral_2007}, which has used a 
microscopic description for the interfacial region at the expense 
of having to carry out numerically most of the analysis, the 
theoretical approach so far has been to map these systems onto the 
case of classic phoresis in an \textit{externally} imposed gradient 
of a field such as, e.g., a solute concentration or an electric 
potential (see Ref. \cite{Anderson_1989} for a review of the theory 
of classic phoresis and additional references; a detailed discussion 
of the shortcomings of such a mapping is provided in 
Ref. \cite{Popescu_2009}).

Starting from the model system proposed in Refs.
\cite{Paxton_2004,Golestanian_2005}, here we study the phoretic 
motion of a spheroidal particle which generates number density 
gradients of product molecules emerging from chemical reactions, 
which are active on the surface of the particle in a cap-like region 
centered at one of the poles. The product molecules diffuse into the 
surrounding unbounded three-dimensional Newtonian liquid solvent. 
Similar to the earlier studies in Refs.
\cite{Golestanian_2005,Golestanian_2007,Popescu_2009,Prost_2009}, 
our work is based on adopting the standard theory of phoresis for 
the present case, in which the gradients are self-generated rather 
than being produced and maintained by external sources \cite{note_1}. 
The motivation for this work is to provide a unified description 
(within the standard theory) for the diffusio-phoretic motion of 
objects belonging to an extended class of geometrical shapes relevant 
to experimental studies 
\cite{Paxton_2004,Howse_2007,Leiderer_2008,Ilona_2009}. The 
previously studied spherical and needle-like shaped objects 
are recovered as particular limiting cases. We note that here we 
focus on the case of rigid particles. If the body is actually soft 
and deformable various additional phenomena, such as a transfer 
between translational and rotational motion upon shape changes, may 
occur (see, e.g., Ref. \cite{Ohta_09}).

The outline of the paper is as follows. In Section \ref{Model} we
define the model. Section \ref{sec_diff_phor_velocity} is devoted
to the derivation of the diffusio-phoretic velocity; it includes 
also the computation of the distribution of the product molecules 
which induces the phoretic motion. The results, as well as the 
connections with the previous studies in Refs.
\cite{Paxton_2004,Golestanian_2005,Golestanian_2007,Saidulu_2008}, 
are discussed in Sec. \ref{discuss}. We conclude with a brief 
summary in Section \ref{summary}.

\section{The Model}
\label{Model}
\begin{figure}[!htb]
\includegraphics[width=1. \linewidth]{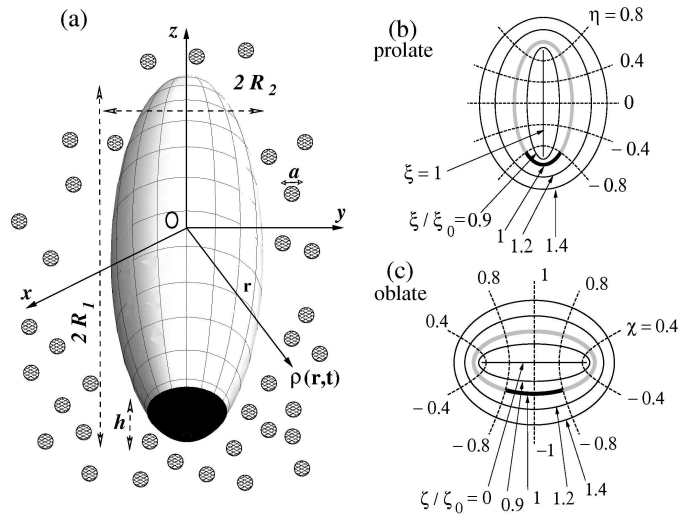}
\caption
{
\label{fig1}
(a) An impermeable, spheroidal particle of polar and equatorial 
semi-axes $R_1$ and $R_2$, respectively, with a cap-like part of 
the surface covered by a catalyst (depicted as a black area). 
The aspect ratio of the particle is $s_r = R_2/R_1$; $s_r < 1$ 
(shown here) corresponds to a prolate shape, while $s_r > 1$ to an 
oblate one. $s_r = 1$ corresponds to a sphere. The product molecules 
of diameter $a$ are shown as small hatched circles. O denotes the 
geometric center of the particle from which the number density 
$\rho(\mathbf{r},t)$ of product molecules is measured. (b) and (c) 
show cuts of the $xz$ plane through the prolate and oblate 
iso-surfaces in terms of prolate $(\xi,\phi,\eta)$ and oblate 
$(\zeta,\phi,\chi)$ spheroidal coordinates, respectively [see, c.f., 
Eqs. (\ref{prolate_coord})-(\ref{def_eta0}) and Eqs. 
(\ref{oblate_coord})-(\ref{def_zeta0}), respectively]. $\xi_0$ and 
$\zeta_0$ denote the values of the prolate $\xi$ and oblate $\zeta$ 
coordinates, respectively, for which the corresponding iso-surfaces 
coincide with the surface of the particle (shown as a thick gray line; 
the thick black line at the lower pole indicates the catalyst covered 
region). 
}
\end{figure}

The system we consider is shown in Fig.~\ref{fig1}(a). It consists of
an impermeable, spheroidal, rigid particle of polar and 
equatorial semi-axes $R_1$ and $R_2$, respectively. At one of the 
poles there is a cap-like region (the black area in Fig.~\ref{fig1}) 
covered by a catalyst [with density $\sigma$ (number of catalytically 
active sites/unit area)] which promotes the chemical conversion of a 
surrounding solvent (not shown in Fig. 1) into product molecules of 
diameter $a$ [small hatched circles in Fig.~\ref{fig1}(a)]. (This is 
a so-called ``Janus particle''\cite{Golestanian_2007}.) 
%%%%%%%%%%%%%%%%%%%%%%%%%
%%%%%%%%%%%%%%%%%%%%%%%%%

\begin{figure}[!htb]
\includegraphics[width= .9 \linewidth]{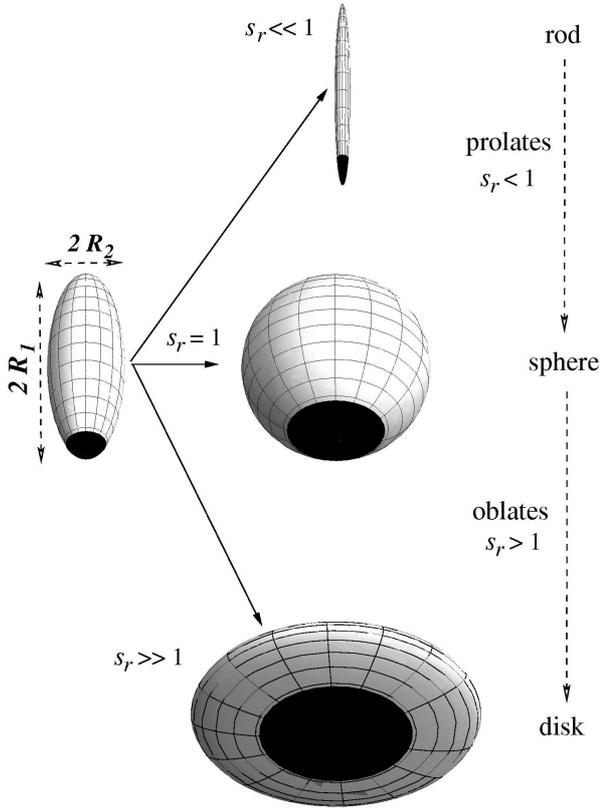}
\caption
{
\label{fig2}
A schematic representation of the spheroidal shapes and of the 
limiting cases which are discussed in the main text; $s_r = R_2/R_1$ 
is varied by varying $R_2$ while keeping $R_1$ fixed.
}
\end{figure}

The shape of the object is characterized by its aspect ratio 
$s_r = R_2/R_1$: $s_r < 1$ refers to a prolate spheroid, 
$s_r > 1$ refers to an oblate spheroid, and $s_r = 1$ to 
a sphere (see Fig. \ref{fig2}). 
%%%%%%%%%%%%%%%%%%%%%%%%%
%%%%%%%%%%%%%%%%%%%%%%%%%
The various spheroidal shapes, i.e., the values of $s_r$ in the full 
range $0 < s_r < \infty$, are systematically explored by considering 
the parameter $R_1$ fixed (introducing a characteristic length scale) 
while varying the parameter $R_2$. The cartesian coordinate system is 
chosen such that for prolates the foci of the generating (through 
rotation around its \textit{major} axis) ellipse are located on the 
$z$ axis symmetrically with respect to the origin O which is located 
at the geometrical center of the particle; correspondingly, for 
oblates the foci of the generating (through rotation around its 
\textit{minor} axis) ellipse are located in the $xy$ plane 
symmetrically with respect to the origin O. The ratio $s_h = h/R_1$ 
between the height of the cap-like catalyst covered area and the 
polar semi-axis characterizes the fraction of the particle surface 
covered by catalyst. Note that the case of a particle with a 
point-like catalytic site located at one of its poles can be included 
in the model by considering the limit 
$\{s_h \to 0,~\sigma \to \infty \}$ under the constraint that 
$\sigma \times s_h$ is finite (see, c.f., Appendix \ref{app_B}). 
Therefore, by varying the parameters $s_r$ and $s_h$ the geometries 
studied in the literature are recovered as the following limiting 
cases (see also Ref. \cite[(a)]{Happel_book}):
\begin{itemize}
\item $\{s_r \to 1^-, 0 < s_h < 1\}$  (where $1^-$ indicates that 
the limit is taken through prolate shapes, i.e., $s_r \lesssim 1$) 
corresponds to a sphere partially covered by catalyst as in 
Ref. \cite{Golestanian_2007}. (Note that it can be similarly 
obtained through oblate shapes, i.e., $s_r \gtrsim 1$, as the 
limiting case $\{s_r \to 1^+, 0 < s_h < 1\}$);
\item $\{s_r \to 1^-, s_h \to 0\}$ (or, equivalently, 
$\{s_r \to 1^+, s_h \to 0\}$) with $\sigma s_h$ finite 
corresponds to a sphere (obtained as the limit of prolate or 
oblate shapes, respectively) with a point-like catalytic site as 
in Ref. \cite{Golestanian_2005}; 
\item $\{s_r \ll 1,~0 < s_h < 1\}$ corresponds to the limit of a 
needle-like particle, which approximates an elongated cylinder (rod) 
partially covered by catalyst as considered in Refs. 
\cite{Paxton_2004,Saidulu_2008,Golestanian_2007}.
\end{itemize}
Additionally, 
\begin{itemize}
\item $\{s_r \gg 1,~0 < s_h < 1\}$ approximates a thin disk 
partially covered by catalyst.
\end{itemize}

In general, the chemical conversion of a solvent generates several 
types of product molecules. Here we shall focus on the particular 
case in which the chemical conversion 
$A \stackrel{\text{cat}}{\to} A' + B$ of a solvent molecule ($A$) 
leads to two molecules ($A'$ and $B$) only, one very similar in size 
and properties with the solvent itself ($A' \approx A$), the other 
one ($B$) significantly different. In the following only this latter 
is denoted as ``product molecule'' and plays the role of a solute in 
the solvent. In other words, we consider a situation in which the net 
result of the chemical conversion can be approximated as the 
generation of a solute, solely, and in which the reaction does not 
lead to a solvent depletion near the catalytic site (which otherwise 
would act as a solvent sink). For example, this is approximately the 
case for the Pt catalyzed decomposition of hydrogen peroxide 
(H$_2$O$_2$) in aqueous solution into water (H$_2$O) and oxygen 
(O$_2$) molecules, as discussed in 
Refs. \cite{Paxton_2005,Howse_2007}. In these experimental studies
the oxygen plays the role of the product molecule the properties of
which differ significantly from those of the solvent. Here the
solvent is actually a binary liquid mixture of H$_2$O and H$_2$O$_2$
for which H$_2$O is chemically passive and does not participate in 
the chemical conversion.

We thus assume that the reaction at the catalytic zone, i.e., the
cap-like area centered around the pole at $- R_1  \mathbf{\hat e}_z$
(where $\mathbf{\hat e}_z$ is the unit vector of the $z$-axis), acts
effectively only as an ensemble of independent sources -- uniformly 
distributed over the cap area [with number density $\sigma$ 
(number of catalytically active sites/unit area)] -- of product 
molecules of diameter $a$, which are diffusing into the solvent with 
diffusion coefficient $D$ \cite{Golestanian_2007}. We shall focus on 
the case in which the reaction rate $\nu_B$ at a catalytic
site, i.e., the number of product molecules created per unit time,
is independent of time. The number density of product molecules is
considered to be so low that among themselves they behave like an
ideal gas. There is an interaction potential between the product
molecules and the moving particle, which includes the impermeability
condition at the surface of particle. The interactions between the
product molecules and the solvent are accounted for in an effective
way via the Stokes - Einstein expression $D = k_B T/(3 \pi \mu a)$
for the diffusion coefficient $D$ of the product molecules
\cite{S-E relation}, where $k_B$ is the Boltzmann constant, $T$ is
the temperature, and $\mu$ is the viscosity of the solution (solvent
plus the solute, i.e., the product molecules).
\section{Phoretic velocity}
\label{sec_diff_phor_velocity}

\subsection{General considerations}
\label{phor_vel_general_subsec}
The presence of a source of solute (product molecules) on parts of
the surface of the particle creates a non-uniform and time dependent
distribution of solute in the solution (see Fig. \ref{fig1}), i.e., a
non-uniform composition of the solution. The solute number density
$\rho(\mathbf{r},t)$ is characterized by two important features. On 
the scale of the particle size $\min(R_1,R_2) \gg a$, $\rho$ varies 
due to the diffusion process. On the much smaller length scale of 
the solute diameter $a$, $\rho(\mathbf{r},t)$ varies also near the 
particle surface because it interacts with the particle via an 
\textit{effective} substrate potential $\Psi$ (in the sense that 
$\Psi$ describes the interaction of the solute molecule with the 
particle in the presence of the solvent). Typically the range 
$\delta$ of $\Psi$ is proportional to the solute diameter $a$. 
Accordingly the solute molecules interact directly with the particle 
only if they are within a thin surface film of thickness $\delta$, 
which is assumed to be not deformed by the motion of the particle 
\cite{Anderson_1989}.

In what follows, we assume that the solvent can be considered as a
continuum with constant density $\rho_{solv}$ even at the length
scale $\delta$, such that a hydrodynamic description applies within
the aforementioned surface film. Within this picture the solute
molecules and their effective interaction with the particle are
replaced by a corresponding distribution of ``point forces'' acting
on the solvent in the film. Within the limitations of such an
approach the hydrodynamic description of the solution naturally
splits into that of an ``inner'' region formed by the surface film
and that of an ``outer'' region formed by the exterior space beyond
the surface film. Furthermore we assume that for typical velocities 
$V$ in phoresis both the Reynolds number $\mathrm{Re} \simeq 
\tilde \rho_{solv} V \max(R_1, R_2)/\mu$, where $\tilde \rho_{solv}$ 
is the \textit{mass density} of the solvent,
and the Peclet number $\mathrm{Pe} \simeq V \max(R_1, R_2) /D$ are
small, such that one can approximate the hydrodynamic description
with the creeping flow (Stokes) equations and disregard the
convection of the solute compared to its diffusive transport. (Here
we have assumed that the magnitude of the hydrodynamic flow velocity
$\mathbf{u}$ is similar to that of the phoretic velocity $V$, an
assumption which will be justified \textit{a posteriori}.) For a
particle of size $\max(R_1,R_2) = 10~\mu$m moving through water
(density $\tilde \rho = 10^3~\mathrm{kg/m}^3$, viscosity $\mu =
10^{-3}~\mathrm{Pa~s}$) with a velocity of the order of $\mu$m/s,
which is typical for phoresis, one has $\mathrm{Re} \simeq 10^{-5}$.
For the diffusion at room temperature ($k_B T_{room} \sim 4 \times
10^{-21}$ J) of O$_2$ ($a \sim 10^{-10}$ m) in an aqueous H$_2$O$_2$ 
solution ($\mu \simeq 10^{-3}~\mathrm{Pa~s}$), the Stokes-Einstein 
relation leads to an estimate 
$D \simeq 4 \times 10^{-9}~\mathrm{m}^2/\mathrm{s}$ for the diffusion 
coefficient (in agreement with Ref. \cite{Paxton_review_2006}),
and thus $\mathrm{Pe} \simeq 10^{-2}$. Therefore both latter 
assumptions are justified.

Following Refs.
\cite{Anderson_1989,Ajdari_2006,Golestanian_2007}, the ensuing
asymmetric, non-uniform solute number density $\rho(\mathbf{r},t)$
around the  particle will give rise, \textit{within the surface film
only}, to a pressure gradient along the surface of the particle. This
is the case because the surface film is very thin on the scale of the 
particle size $\min(R_1,R_2)$ so that the equilibration of the 
composition profile of the solution within the surface film can be 
assumed to be fast along the direction normal to the surface of the 
particle compared with the diffusional relaxation time of the 
composition gradient along the surface of the particle, which 
typically involves a length scale of the order of $\max(R_1,R_2)$. 
In accordance with our earlier assumption that the solute particles 
can be considered to form an ideal gas, this implies that near the 
surface of the particle the spatial variations of $\rho$ (within the 
surface film and in the direction normal to the surface of the 
particle) are given by a Boltzmann distribution $\exp( - \beta \Psi)$, 
where $\beta = 1/(k_B T$), corresponding to the local equilibrium 
configuration in the presence of the effective interaction potential 
between the particle and the solute molecules, with a prefactor which 
depends on the position along the surface of the particle 
\cite{Anderson_1989}. Mechanical equilibrium of the solvent within 
the surface film along the direction normal to the surface of the 
particle (no flow along this direction) requires the pressure gradient 
along the normal to be equal to the body force densities due to the 
effective particle-solute interactions. Therefore the pressure within 
the surface film differs from the ``outer'' pressure field by an
``osmotic pressure'' term, i.e., a term proportional to the extra
solute density $\Delta \rho$ in excess to that density in the outer 
region. Since this osmotic pressure varies along the surface of the
particle, there is a pressure gradient along the surface of
the particle. This lateral pressure gradient is not balanced by
any body force and thus generates shear stress within the surface
film. Therefore it induces hydrodynamic flow of the solution along
the surface of the particle and entails motion of the particle with
a steady-state velocity $\mathbf{V}$ \cite{note_2,Morrison_1970}.
Because the system has azimuthal symmetry, the motion is along the 
$z$-axis, i.e., $\mathbf{V} = V \mathbf{\hat e}_z$.

We note that in a number of instances a different argument has been 
used in deriving an expression for the velocity of the particle. It 
invokes a balance between a Stokes-like viscous friction and a 
``driving force'' produced, e.g., by a spatially non-uniform surface 
tension (owing to the product molecules changing the composition of 
the solution) \cite{Paxton_2004,Saidulu_2008}. Such an argument (see, 
e.g., Ref. \cite{Paxton_2004} and the follow-up 
Ref. \cite{Saidulu_2008} which aimed at improving
the description through a more rigorous calculation of the density
of product particles) is in sharp contradiction with the generally
accepted view that in phoresis the motion of the particle plus the
surface film (i.e., the region within which the interaction between 
the particle and the solute molecules is relevant) is force-free, 
with the velocity being determined precisely by this requirement 
\cite{Anderson_1989,Golestanian_2007,Popescu_2009,Prost_2009}. This 
confusion leads to results which differ from the correct ones roughly 
by a factor which is the square of the ratio between the size of the 
particle and the range of the particle-solute interactions; i.e., the 
predicted values of the velocity are too large by orders of magnitude.

\subsection{Phoretic slip and phoretic velocity}
\label{phor_slip_and_vel_subsec}
As discussed above, the pressure gradient along the surface of the
particle, induced by its interaction with the non-uniform
distribution $\rho(\mathbf{r},t)$ of product molecules within a thin 
surface film, leads to flow of the solution relative to the particle. 
Assuming that $\rho(\mathbf{r},t)$ is changing slowly in time, this 
hydrodynamic flow, considered to be locally planar, translates 
into a (phoretic) slip-velocity,
\begin{equation}
\label{slip_vel}
\mathbf{v}_s (\mathbf{r}_p) = - b \nabla^\Sigma
\rho(\mathbf{r}_p)\,,\textrm{ for } \mathbf{r}_p \in \Sigma_\delta\,,
\end{equation}
as a boundary condition for the hydrodynamic flow in the outer
region \cite{Anderson_1989,Golestanian_2007,Popescu_2009}. In this 
equation $\Sigma_\delta$ denotes the outer edge of the surface film 
(which is a surface at a distance $\delta$ parallel to the surface 
$\Sigma$ of the particle), $\mathbf{r}_p$ denotes a point $P$ 
on $\Sigma_\delta$, $\nabla^\Sigma$ denotes the projection of the 
gradient operator onto the corresponding local tangential plane of 
the surface of the  particle (actually of $\Sigma_\delta$; but for 
the outer problem this can be replaced by $\Sigma$ because the 
variations of $\rho$ are over length scales much larger then 
$\delta$), while
\begin{equation}
\label{b_def}
b = \dfrac{k_B T}{\mu} \Lambda
\end{equation}
is an effective ``mobility'' coefficient, and $\lambda =
\sqrt{|\Lambda|}$ is a characteristic length scale.
The latter is given in terms of the effective interaction potential
$\Psi$ between the particle and the product molecules. Within a 
local coordinate system in a small domain of the surface film of 
width $\delta$, centered at position $\mathbf{r}_p$, $\Psi$ 
determines the product molecule distribution along the direction 
$\hat y$ normal to the particle surface and yields 
\cite{Anderson_1989,Popescu_2009}
\begin{equation}
\label{lambda_def}
\Lambda \equiv \Lambda(\mathbf{r}_p) =
\int\limits_0^\infty \, d \hat y\, \hat y
\left(e^{-\beta \Psi(\hat y;\,\mathbf{r}_p)} - 1\right)\,.
\end{equation}

We have explicitly indicated in Eq. (\ref{lambda_def}) that the
effective interaction potential $\Psi$ and, consequently, the 
length scale $\lambda$ may vary slowly along the surface of the 
particle over length scales much larger than the thickness $\delta$ 
of the surface film. This is the case because one may reasonably 
expect that the effective interactions between the particle and 
the product molecules depends on the local chemical composition of 
the surface. Thus they can be different in the region $\Sigma_c$ at 
the pole which is covered by the catalyst from that at the 
chemically inert part of the particle surface. We note that such 
non-uniformity of the surface properties has been also explicitly 
noted in, e.g., Ref. \cite{Paxton_2004}, anticipating that the 
surface tension of the rod-solution interface is different at the 
Pt and Au ends of the rod. However, its role was ignored in the 
analysis there, as well as in the follow-up work in Ref. 
\cite{Saidulu_2008}, based on the argument that the density of O$_2$ 
is uniform over the Pt end. Nevertheless, such an argument is clearly 
contradicted by the theoretically calculated O$_2$ density along the 
surface of the rod which shows significant variations over the whole 
surface of the rod (see Fig. 4 in Ref. \cite{Saidulu_2008}). In a 
first order approximation one can account for these non-uniform 
properties along the surface by using a position dependent effective 
mobility $b(\mathbf{r}_p)$ [Eq. (\ref{b_def}) with $\Lambda \to 
\Lambda(\mathbf{r}_p)$], which can attain two values, in the 
expression of the phoretic slip-velocity [Eq. (\ref{slip_vel})]. 
This approach has been pursued in Ref. \cite{Golestanian_2007}. 
In the case of a sharp boundary between the catalyst-covered region 
and the inert one such an approximation is probably justified. However, 
if the non-uniform properties vary smoothly and significantly over 
the surface, then it becomes unclear if this can be accounted for 
solely by a piece-wise constant effective mobility $b(\mathbf{r}_p)$. 
In such cases, a detailed analysis of the hydrodynamic flow in the 
surface film forming the inner region is required for understanding 
the effects of inhomogeneities on the induced phoretic motion. For 
example, it has been shown that a solid sphere with position-dependent 
slip boundary condition immersed in a laminar flow experiences a 
torque induced by this inhomogeneity \cite{Willmott_09}. 
Moreover, in the case of electrophoresis it has already been shown 
that for a particle with non-uniform surface properties the phoretic 
motion depends strongly on the details of this non-uniformity, which 
may give rise to counter-intuitive results such as electrophoresis 
of spherical particles which are electrically neutral and have a 
zero mean $\zeta$-potential (see, e.g., 
Refs. \cite{Anderson_1984,Fair_1988}). 
But there may be cases in which such coatings by certain catalysts 
do not significantly change the effective interaction between the 
particle and the product molecules. Accordingly, here we shall focus 
on the case in which $\Psi$, and thus the effective mobility $b$, 
can be assumed to be constant over the surface of the particle. The 
general case of an effective interaction potential which varies 
over the surface of the particle will be discussed elsewhere.

As mentioned above, the dynamics at small Pe numbers implies that 
the convection of the solute is negligible compared to the diffusive 
transport, i.e., the diffusion of the solute is apparently decoupled 
from the hydrodynamic flow. However, a coupling between the density 
profile of the solute, i.e., the solute diffusion, and the flow of 
the solution is re-established by Eq. (\ref{slip_vel}) which, in the 
reference frame co-moving with the particle, represents the boundary 
condition (BC) at the edge of the surface film for hydrodynamic flow 
in the outer region. Because there are no forces acting on the 
solution beyond the surface film, based on the assumption of low Re 
numbers the hydrodynamic flow in the outer region is obtained as the 
solution of force free and incompressible Stokes equations subject 
to the following BCs in the laboratory frame: (i) prescribed velocity 
$\mathbf{V} + \mathbf{v}_s$ at the edge $\Sigma_\delta \simeq \Sigma$ 
of the surface film (i.e., sticking on the surface of the particle 
plus a slip velocity $\mathbf{v}_s$ at the edge of the surface film) 
and (ii) zero velocity (fluid at rest) far away from the particle.
After computing the solution in the outer region, which depends
parametrically on the velocity $\mathbf{V}$ of the particle via the 
above BC (i), the phoretic velocity is determined by the condition 
that the motion of the system composed of the particle plus the 
surface film is force free. Here the cumbersome explicit calculation 
of the hydrodynamic flow is avoided by using Brenner's generalized 
reciprocal theorem \cite[(b)]{Happel_book}, which allows one to express 
the phoretic velocity of the particle as a surface integral of the
phoretic slip velocity weighted by the normal component of the 
position vector (see Appendix A and, e.g., Ref. \cite{Fair_1988}):
\begin{eqnarray}
\label{velocity}
\mathbf{V} &=& - \dfrac{1}{3\,{\cal V}_p} \iint \limits_{\Sigma_\delta}
(\mathbf{\hat n} \, \cdot \mathbf{r}_p) 
\,\mathbf{v}_s \,d\,\Sigma_\delta \,,\nonumber\\
&& ~\nonumber\\
V &\simeq&
\dfrac{b}{3\,{\cal V}_p} \iint \limits_{\Sigma}
(\mathbf{\hat n} \cdot \mathbf{r}_p) 
\,(\mathbf{\hat e}_z \cdot \nabla^\Sigma \rho)\,\,d\,\Sigma\,,
\end{eqnarray}
where the second equality follows by using Eq. (\ref{slip_vel}) to 
replace $\mathbf{v}_s$.

In this equation ${\cal V}_p = (4/3) \pi s_r^2 R_1^3$ denotes the
volume of the particle, $\mathbf{\hat n}$ is the unit vector of the
direction normal to the surface of the particle, and in the integral
we replaced $\Sigma_\delta$ by $\Sigma$ because the flow in the outer
region varies over length scales which are much larger than the
difference $\delta$ between the two. According to Eq.
(\ref{velocity}), knowledge of the number density $\rho$ of product
molecules at the outer edge $\Sigma_\delta$ of the surface film
completely determines the velocity $\mathbf{V}$ of the particle.

\subsection{Steady-state distribution of product molecules (solute)}
\label{solute_distr_subsec}

After switching on the catalytic reaction there is an initial 
transient phase during which the distribution $\rho(\mathbf{r},t)$ of 
product molecules builds up. Because we are dealing with an unbounded 
solvent, one may expect that at long times ($t \to \infty$) the 
distribution of product molecules reaches a steady-state distribution 
$\rho(\mathbf{r})$. In what follows, we focus on the motion in the 
long-time regime (in practice, at times $t$ much longer than the 
diffusion time $t_D = [\max(R_1,R_2)]^2/D$; for the numerical 
estimates from Subsec. \ref{phor_vel_general_subsec}, 
$t_D \simeq 2.5$ s) in which $\rho$ and consequently the phoretic 
slip ${\mathbf v}_s$ and the phoretic velocity $\mathbf V$ [see 
Eqs. (\ref{slip_vel}) and (\ref{velocity}), respectively] become 
time independent. Note that here we implicitly assume 
that the time-scale $t_R$ of the rotational diffusion of the polar 
axis of the particle is sufficiently larger than $t_D$ such that 
the steady-state distribution $\rho(\mathbf{r})$ is attained and 
translational motion of the particle with uniform velocity $V$ 
occurs; at times $t \gg t_R$ the rotational diffusion of the polar 
axis leads to a quasi-diffusive behavior of the particle displacement 
(see, e.g., Refs.~\cite{Howse_2007,Golestanian_2009}). These two 
regimes are clearly observable in the experiments discussed in 
Ref. \cite{Paxton_2004} (see, in particular, the supporting 
information therein).

Within the assumptions that the diffusion of product molecules is
fast compared with the convection by the solvent flow, i.e., in the
limit of small Peclet numbers, and that the product distribution
$\rho(\mathbf{r})$ in the steady state is undisturbed by the flow,
i.e., neglecting any so-called polarization effects of the surface
film \cite{Anderson_1989}, the steady state distribution
$\rho(\mathbf{r})$ of product molecules in the outer region around
the moving particle is governed, in the co-moving frame, by the
diffusion equation
\begin{equation}
\label{diff_eq}
D \nabla^2 \rho(\mathbf{r}) = 0, \,\mathbf{r} \in
\text{outer region}\,.
\end{equation}
This equation is to be solved subject to the BCs of (i) zero density 
far away from the outer edge $\Sigma_\delta$ of the surface film and 
(ii) at each point $\mathbf{r} \in \Sigma_\delta$ the product 
molecules current in the outward direction $\mathbf{\hat n}$ normal 
to the surface is equal to the total reaction rate in an infinitesimal 
element of $\Sigma$ centered at $\mathbf{r} - \delta \mathbf{\hat n}$. 
The latter BC holds within the assumption that the surface film is 
very thin [$\delta \ll \min(R_1,R_2)$] such that in the steady state 
the lateral transport of product molecules in the surface film is 
negligible compared to the transport into the outer region along the 
direction normal to $\Sigma_\delta$. Since in the calculation of 
$\rho(\mathbf{r})$ we shall replace the values of the coordinates 
corresponding to $\Sigma_\delta$ by those of $\Sigma$, justified by 
the same argument that $\rho(\mathbf{r})$ is expected to vary over 
length scales which are much larger than $\delta$, we shall formally 
apply the second boundary condition directly on the surface $\Sigma$ 
of the particle, rather than on $\Sigma_\delta$. Hence, the BCs take 
the following forms:
\begin{subequations}
\label{BC}
\begin{equation}
\label{farfield_BC}
\rho(|\mathbf{r}| \to \infty) = 0\,
\end{equation}
\begin{eqnarray}
\label{normalJ_BC}
&& - D \left.\left[\mathbf{\hat n} \cdot \nabla \rho(\mathbf{r})
\right]\right|_{\mathbf{r} \in \Sigma} = 
\nu_B \,\sigma\, \Upsilon(\mathbf{r})\hspace*{1.cm} \\
&& \hspace*{2.5cm} =:
\nu_B \,\sigma\,\,
\begin{cases}
1\,,~
\mathbf{r} \in  \mathrm{catalytic~cap}\,,\\
0\,,~ \mathrm{otherwise}\,.\hspace*{1cm}
\end{cases}\nonumber
\end{eqnarray}
\end{subequations}

\begin{figure}[!htb]
\begin{center}
\includegraphics[width=.7 \linewidth]{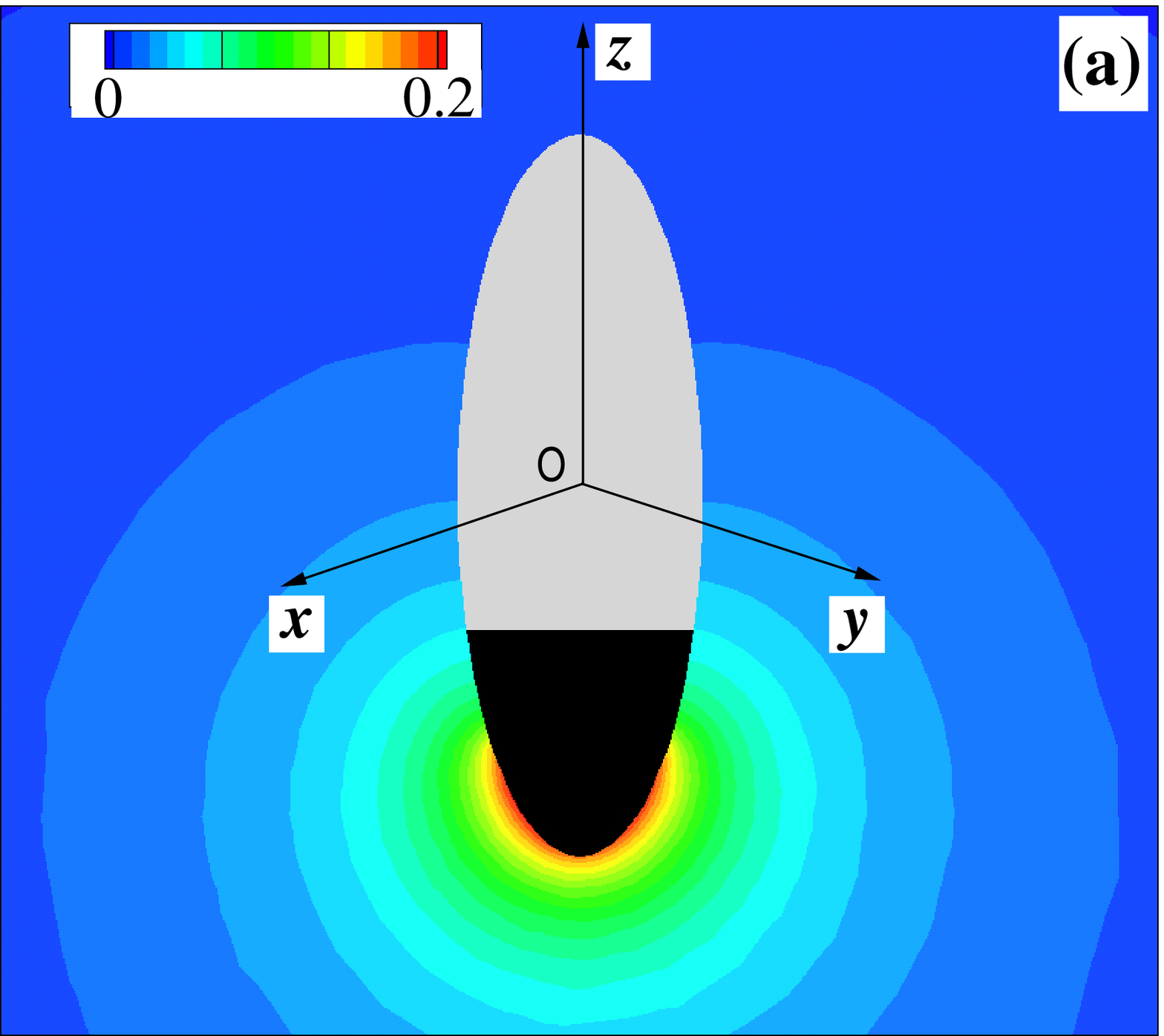}\\
\vspace*{.1cm}
\includegraphics[width=.7 \linewidth]{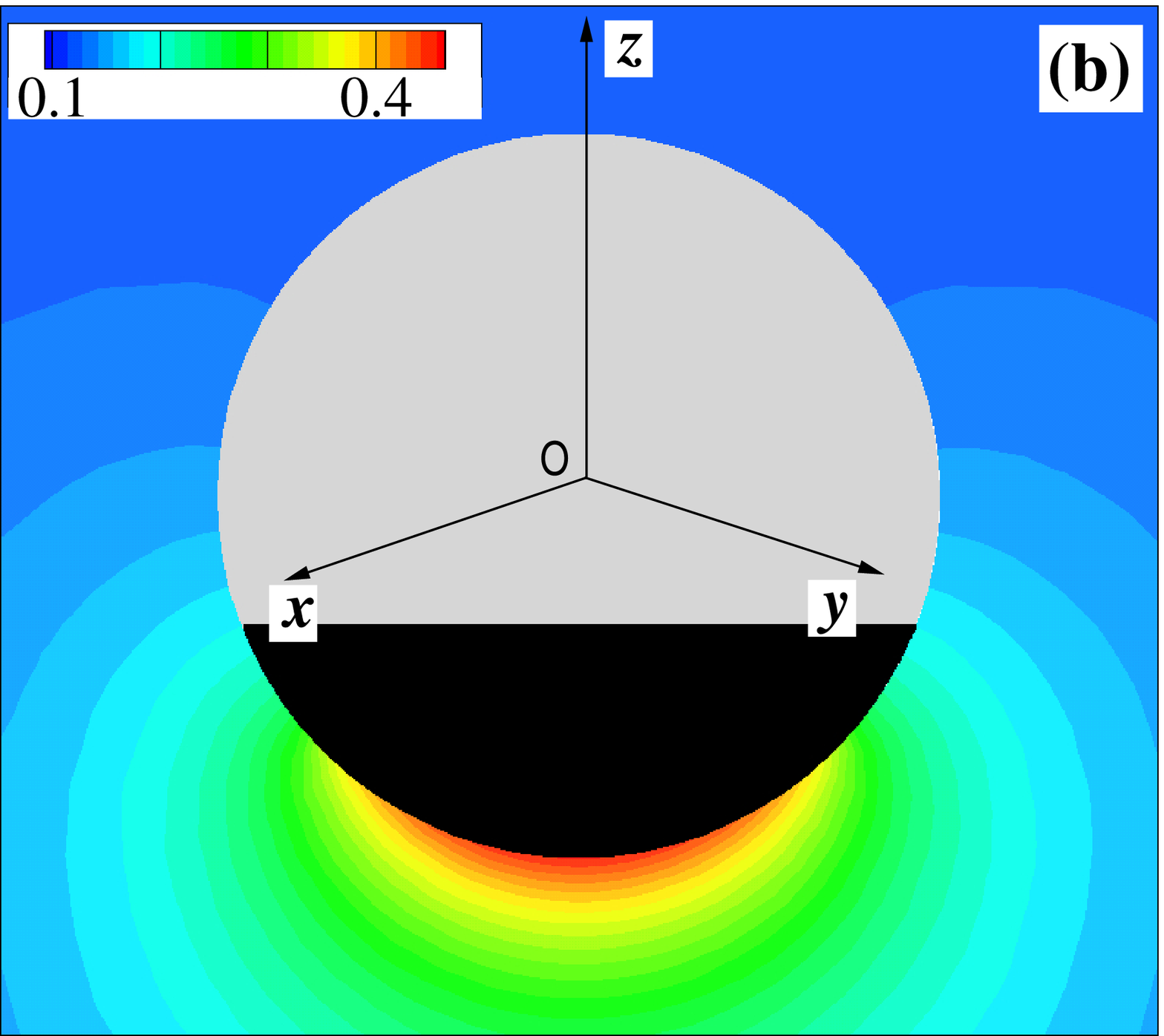}\\
\vspace*{.1cm}
\includegraphics[width=.7 \linewidth]{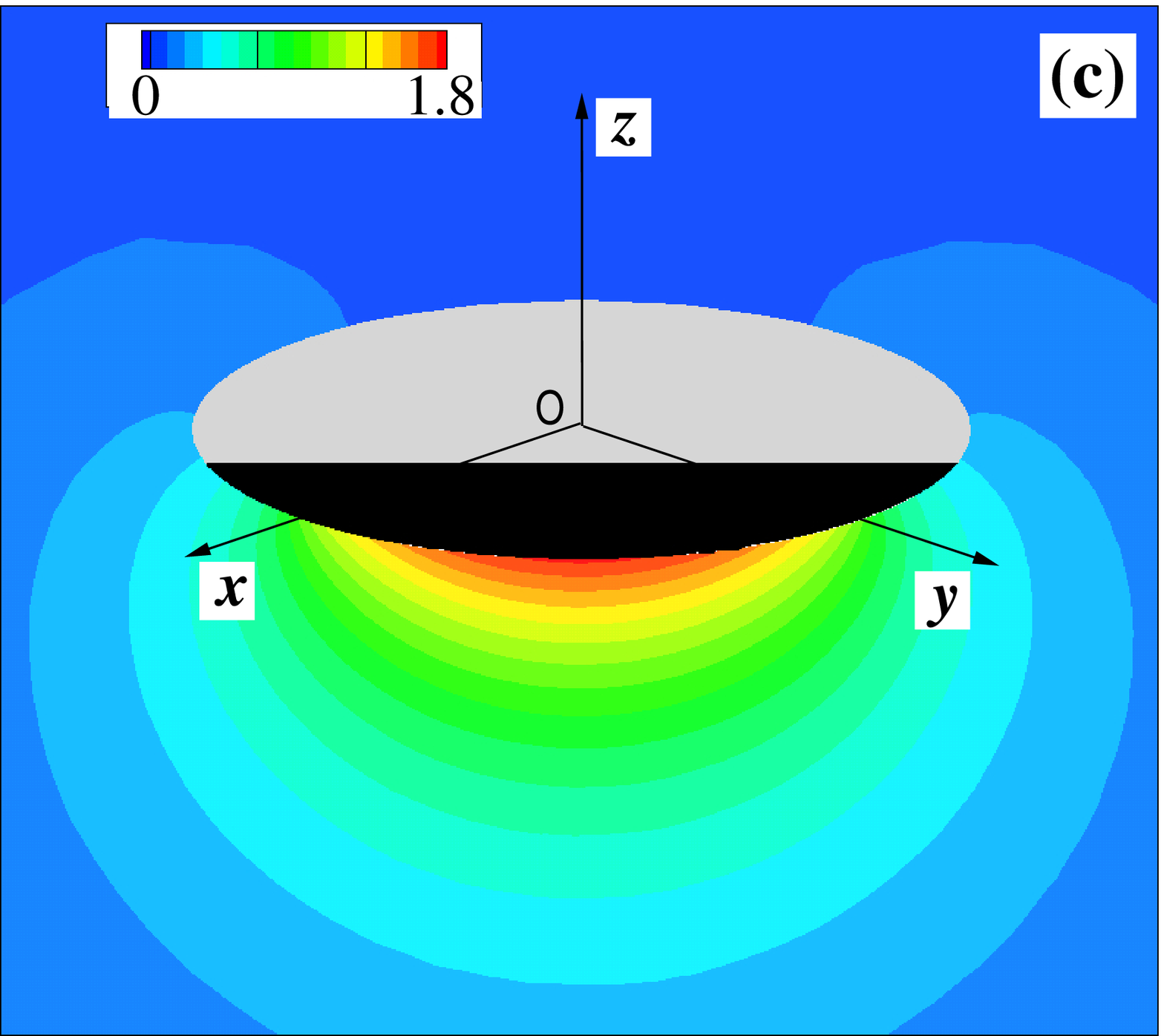}
\end{center}
\caption
{
\label{fig3}
Steady-state distributions $\rho(\mathbf{r})$ of the solute 
(in units of $\rho_0 = \nu_B \,\sigma\,R_1/D$) around (a) a prolate 
($s_r = 0.3$), (b) a spherical ($s_r = 1$), and (c) an oblate 
($s_r = 3.0$) particle partially covered by catalyst ($s_h = 0.3$). 
These distributions are obtained by numerical integration of 
Eqs. (\ref{diff_eq}) and (\ref{BC}). In each case the color coding 
of the density ranges from blue (low density) to red (high density) 
in accordance with the corresponding scale shown on the top left. 
Note the differences in the scales for (a), (b), and (c).
}
\end{figure}

Examples of steady-state distributions $\rho(\mathbf{r})$ of the 
solute (in units of $\rho_0 = \nu_B \,\sigma\,R_1/D$), obtained 
from a direct numerical integration of Eqs. (\ref{diff_eq}) and 
(\ref{BC}), are shown in Fig. \ref{fig3}.
%%%%%%%%%%%%%%%%%%%%%%%%%
%%%%%%%%%%%%%%%%%%%%%%%%% 
One notes that as expected (see Subsec. \ref{phor_vel_general_subsec}) 
there are significant density variations over the part of the surface 
covered by the catalyst. These are more pronounced (in addition 
to the larger amplitude of the distribution) for the ``flatter`` 
shapes, thus intuitively suggesting a larger resulting phoretic 
velocity for the more flatter shapes. In the following subsections 
we derive the solute distributions analytically for any spheroidal 
shape, which allows us to discuss quantitatively the effect of the 
shape on the phoretic velocity.

\subsubsection{Prolate particles}
\label{solute_distr_prolates}
For a prolate shaped ($s_r < 1$) particle with the catalyst 
distributed over a cap-like region centered at one of its poles 
(see Fig. \ref{fig1}), Eq. (\ref{diff_eq}) subject to the BCs given by 
Eq. (\ref{BC}) is most conveniently solved in terms of the prolate 
spheroidal coordinates $(\xi,\phi,\eta)$ 
(see Ref. \cite[(a)]{Abramowitz_book}):
\begin{eqnarray}
\label{prolate_coord}
x &=& \kappa \sqrt{(\xi^2-1)(1-\eta^2)} \cos \phi \,,\\
y &=& \kappa \sqrt{(\xi^2-1)(1-\eta^2)} \sin \phi \,,\\
z &=& \kappa \xi \eta \,.
\end{eqnarray}
In these equations $\xi \geq 1$ and $-1 \leq \eta \leq 1$ parameterize 
confocal ellipsoids and hyperboloids of revolution, respectively, which 
have their foci placed on the $z$ axis a distance $2 \kappa$ apart and 
symmetrically with respect to the origin O, while $0 \leq \phi < 2\pi$ 
is the azimuthal angle. The $xz$ planar cuts through the iso-surfaces 
in terms of prolate spheroidal coordinates are shown in 
Fig. \ref{fig1}(b). 
The choice
\begin{equation}
\label{parameter_k}
\kappa = \kappa_p = \sqrt{R_1^2 - R_2^2} = R_1 \sqrt{1-s_r^2}\,,
\end{equation}
ensures that the family of ellipsoids includes the one,
\begin{equation}
\label{def_xi0}
\xi_0 = \dfrac{1}{\sqrt{1-s_r^2}} > 1\,,
\end{equation}
which corresponds to the surface of the particle [$\xi_0$ follows
from Eqs. (\ref{prolate_coord})-(\ref{parameter_k})
and $z(\xi_0,\eta = \pm 1; \kappa_p) = R_1$]. The outer region
thus corresponds to $\xi > \xi_0$. Since the intersection of any
hyperboloid $\eta = const$ from the family defined by $\kappa_p$ with
the surface of the particle is a circle parallel to the equatorial 
$xy$ plane, the cap-like region on $\Sigma$ covered by the catalyst 
is parameterized by
$(\xi = \xi_0, 0 \leq \phi < 2\pi, -1 \leq \eta \leq \eta_0)$, where
\begin{equation}
\label{def_eta0}
-1 \leq \eta_0 = -1 + s_h \leq 1
\end{equation}
[which follows from Eqs. (\ref{prolate_coord})-(\ref{def_xi0}) and
$z(\xi_0,\eta_0; \kappa_p) = -R_1 + h$, see Fig. \ref{fig1}(a)].

In terms of prolate spheroidal coordinates the solution 
$\rho(\xi,\eta)$ of Eq. (\ref{diff_eq}), which has no dependence on 
$\phi$ due to the azimuthal symmetry of the system, is finite at 
$\eta = \pm 1$ (i.e., on the $z$ axis), and satisfies the BC in 
Eq. (\ref{farfield_BC}), can be written as \cite{Bauer_89,Smythe_68}
\begin{equation}
\label{series_rho_prolate}
\rho(\xi,\eta) = \sum_{\ell \geq 0} 
c_\ell Q_\ell(\xi) P_\ell(\eta)\,.
\end{equation}
$P_\ell$ and $Q_\ell \equiv Q_\ell^{(0)}$ with 
\cite{Smythe_68,Hobson_book}
\begin{eqnarray}
\label{Ql_def}
Q_\ell (w) &:=& 2^\ell \sum_{s \geq 0} 
\dfrac{(\ell+2s)!\,(\ell+s)!}{s! (2 \ell+2s+1)!}w^{-\ell-2s-1}
\nonumber\\
&=& \dfrac{1}{2}\,P_\ell(w)\,\ln\,\dfrac{w+1}{w-1} 
- \dfrac{2 \ell-1}{\ell}\,P_{\ell-1}(w)\,\\
&-& \dfrac{2 \ell-5}{3 (\ell-1)}\,P_{\ell-3}(w)\,-\cdots
\,,~w \in \mathbb{C} \setminus [-1,1]\,,\nonumber
\end{eqnarray}
are the Legendre polynomial and the zeroth-order associated Legendre
function of the second kind of degree $\ell$ 
(Ref. \cite[(b)]{Abramowitz_book}), respectively. The terms with 
$P_{\ell < 0}$ are, by definition, identically zero. The coefficients 
$c_\ell$ are determined by the BC in Eq. (\ref{normalJ_BC}). 
By noting that 
$\mathbf{\hat n} = \mathbf{\hat e}_\xi$, where 
$\mathbf{\hat e}_\xi$ is the unit vector corresponding to the $\xi$ 
direction, Eq. (\ref{normalJ_BC}) can be re-written as
\begin{equation}
\label{normalJ_BC_xi}
- \dfrac{D}{h_\xi(\xi_0,\eta)}\,
\left(\dfrac{\partial \rho(\xi,\eta)}
{\partial \xi}\right)_{\xi = \xi_0} = \nu_B\, \sigma\,
\Upsilon(\eta;\eta_0)\,,
\end{equation}
where $h_\xi$ is the scale factor corresponding to
$\mathbf{\hat e}_\xi$ (with similarly defined scale factors $h_\eta$ 
and $h_\phi$) (Ref. \cite[(a)]{Abramowitz_book}):
\begin{subequations}
\label{metric_factors_prolates}
\begin{equation}
\label{h_xi}
h_\xi \equiv {h_\xi(\xi,\eta)} =
\kappa_p \sqrt{\dfrac{\xi^2-\eta^2}{\xi^2-1}}\,,
\end{equation}
\begin{equation}
\label{h_eta}
h_\eta \equiv {h_\eta(\xi,\eta)} =
\kappa_p \sqrt{\dfrac{\xi^2-\eta^2}{1-\eta^2}}\,,
\end{equation}
\begin{equation}
\label{h_phi_prol}
h_\phi \equiv {h_\phi(\xi,\eta)} =
\kappa_p \sqrt{(\xi^2-1)(1-\eta^2)}\,,
\end{equation}
\end{subequations}
and the parametric dependence of the characteristic function 
$\Upsilon(\mathbf{r})$ on $\eta_0$ [Eq. (\ref{normalJ_BC})] is 
indicated explicitly. By using the orthogonality of the Legendre 
polynomials,
\begin{equation}
\label{Legendre_ortho}
\int\limits_{-1}^{1} d\eta\, P_n(\eta)\,P_m(\eta) =
\dfrac{2}{2n+1} \,\delta_{n\,m}\,,
\end{equation}
where $\delta_{n\,m}$ is the Kronecker delta symbol 
(Ref. \cite[(b)]{Abramowitz_book}), and by combining 
Eqs. (\ref{series_rho_prolate})-(\ref{h_xi}), one finds for the 
coefficients $c_\ell$:
\begin{eqnarray}
\label{expr_c_ell}
c_\ell(\xi_0,\eta_0) &=& - (\ell +\dfrac{1}{2})\,
\dfrac{\nu_B \sigma R_1}{D} \,
\dfrac{\gamma_\ell(\xi_0,\eta_0)}
{Q'_\ell(\xi_0) \xi_0 \sqrt{\xi_0^2 - 1}}\nonumber\\
&:=& - \dfrac{\nu_B \sigma R_1}{D} {\tilde c}_\ell(\xi_0,\eta_0) \,,
\end{eqnarray}
where $Q'_\ell(\xi) \equiv dQ_\ell(\xi)/d\xi$ and
\begin{eqnarray}
\label{gamma_ell}
\gamma_\ell(\xi_0,\eta_0) &=&
\int\limits_{-1}^{1} d\eta\,
\sqrt{\xi_0^2 - \eta^2} \,\Upsilon(\eta;\eta_0)
P_\ell(\eta)\nonumber\\
&=& \int\limits_{-1}^{\eta_0} d\eta\,
\sqrt{\xi_0^2 - \eta^2} \,P_\ell(\eta)\,
\end{eqnarray}
so that the coefficients ${\tilde c}_\ell(\xi_0,\eta_0)$ are 
dimensionless.

\subsubsection{Oblate particles}
\label{solute_distr_oblates}

For an oblate shaped ($s_r >1$) particle with the catalyst distributed 
over a cup-like region centered at one of its poles [see 
Figs. \ref{fig1} and \ref{fig2}], Eq. (\ref{diff_eq}) subject to the 
BCs given by Eq. (\ref{BC}) is most conveniently solved in terms of 
the oblate spheroidal coordinates $(\zeta,\phi,\chi)$ 
(see Ref. \cite[(a)]{Abramowitz_book}):
\begin{eqnarray}
\label{oblate_coord}
x &=& \kappa \sqrt{(\zeta^2+1)(1-\chi^2)} \cos \phi \,,\\
y &=& \kappa \sqrt{(\zeta^2+1)(1-\chi^2)} \sin \phi \,,\\
z &=& \kappa \zeta \chi \,\,.
\end{eqnarray}
In these equations $\zeta \geq 0$ and $-1 \leq \chi \leq 1$ 
parameterize confocal ellipsoids and half-hyperboloids of revolution,
respectively, which have foci placed symmetrically with respect to 
the origin O on the $x$ axis a distance $2 \kappa$ apart, while
$0 \leq \phi < 2\pi$ is the azimuthal angle. Cuts by the $xz$ 
plane through the iso-surfaces in terms of oblate spheroidal 
coordinates are shown in Fig. \ref{fig1}(c). The choice
\begin{equation}
\label{parameter_ko}
\kappa = \kappa_o = \sqrt{R_2^2 - R_1^2} = R_1 \sqrt{s_r^2-1}
\end{equation}
ensures that the family of ellipsoids includes the one,
\begin{equation}
\label{def_zeta0}
\zeta_0 = \dfrac{1}{\sqrt{s_r^2-1}} > 0 \,,
\end{equation}
which corresponds to the surface of the particle; $\zeta_0$ follows
from Eqs. (\ref{oblate_coord})-(\ref{parameter_ko})
and $z(\zeta_0,\chi = \pm 1; \kappa_o) = R_1$. The outer region 
thus corresponds to $\zeta > \zeta_0$. Since the intersection of any
hyperboloid $\chi = const$ from the family characterized by 
$\kappa_o$ with the surface of the particle is a circle parallel to 
the equatorial $x\,y$ plane, the cap-like region on $\Sigma$ covered 
by the catalyst is parameterized by
$(\zeta = \zeta_0, 0 \leq \phi < 2\pi, -1 \leq \chi \leq \chi_0)$, 
where
\begin{equation}
\label{def_chi0}
\chi_0 = -1 + s_h\,.
\end{equation}
This follows from Eqs. (\ref{oblate_coord})-(\ref{def_zeta0}) and
$z(\zeta_0,\chi_0; \kappa_o) = -R_1 + h$ [see Fig. \ref{fig1}(a)].

In oblate spheroidal coordinates the solution $\rho(\zeta,\chi)$ of
Eq. (\ref{diff_eq}), which has no dependence on $\phi$ due to the 
azimuthal symmetry of the system, is finite at $\chi = \pm 1$ (i.e., 
on the $z$ axis), and satisfies the BC in Eq. (\ref{farfield_BC}), 
can be written as \cite{Smythe_68}
\begin{equation}
\label{series_rho_oblate}
\rho(\zeta,\chi) = \sum_{\ell \geq 0} 
m_\ell Q_\ell(\mathrm{i} \zeta) P_\ell(\chi)\,,
\end{equation}
where $\mathrm{i} = \sqrt{-1}$ and [see Eq. (\ref{Ql_def})]
\begin{eqnarray}
\label{Q_imag}
Q_\ell (\mathrm{i} \zeta) 
&:=& \,2^\ell \sum_{s \geq 0} 
\dfrac{(\ell+2s)!\,(\ell+s)!}
{s! (2\ell+2s+1)!}(\mathrm{i}\,\zeta)^{-\ell-2s-1} \nonumber\\
&=&
- \mathrm{i}\,\,(\mathrm{arccot}\, \zeta) P_\ell(\mathrm{i} \zeta) 
- \dfrac{2 \ell-1}{\ell} P_{\ell-1}(\mathrm{i} \zeta)\nonumber\\
&& -
\dfrac{2 \ell-5}{3 (\ell-1)} P_{\ell-3}(\mathrm{i} \zeta)-\cdots~~.
\end{eqnarray}
The second equation, for which we have used 
$(1/2) \ln[(z+1)/(z-1)] = \mathrm{arccoth} (z)$ and 
$\mathrm{arccoth}(\mathrm{i} z) = - \mathrm{i}\, \mathrm{arccot} (z)$, 
emphasizes that $Q_\ell (\mathrm{i} \zeta)$ 
is well defined for all $\zeta \in \mathbb{R}$ \cite{Smythe_68}.
This is needed because, for $s_r > \sqrt{2}$, $\zeta_0$ is smaller 
than 1 [Eq. (\ref{def_zeta0})] and thus the point $z = \mathrm{i}$, 
where the Legendre differential equation is singular, lies inside 
the domain of the solution (in contrast to the case of a prolate, 
for which $\xi_0 > 1$). 
The coefficients $m_\ell$ are determined by the BC in 
Eq. (\ref{normalJ_BC}). Note that because the density 
$\rho(\zeta,\chi) \in \mathbb{R}$, while according to 
Eq. (\ref{Q_imag}) $Q_\ell (\mathrm{i} \zeta)$ is real (imaginary) 
for $\ell$ odd (even), the coefficients $m_\ell$ 
in the series representation, Eq. (\ref{series_rho_oblate}) are 
real (imaginary) for $\ell$ odd (even).

 By using $\mathbf{\hat n} = 
\mathbf{\hat e}_\zeta$, where $\mathbf{\hat e}_\zeta$ is the unit 
vector corresponding to the $\zeta$ direction, Eq. (\ref{normalJ_BC}) 
can be re-written as
\begin{equation}
\label{normalJ_BC_zeta}
- \dfrac{D}{h_\zeta(\zeta_0,\chi)}\,
\left(\dfrac{\partial \rho(\zeta,\chi)}
{\partial \zeta}\right)_{\zeta = \zeta_0} = \nu_B\, \sigma\,
\Upsilon(\chi;\chi_0)\,,
\end{equation}
where $h_\zeta$, $h_\chi$, and $h_\phi$ with
\begin{subequations}
\label{metric_factors_oblates}
\begin{equation}
\label{h_zeta}
h_\zeta \equiv {h_\zeta(\zeta,\chi)} =
\kappa_o \sqrt{\dfrac{\zeta^2+\chi^2}{1+\zeta^2}}\,,
\end{equation}
\begin{equation}
\label{h_chi}
h_\chi \equiv {h_\chi(\zeta,\chi)} =
\kappa_o \sqrt{\dfrac{\zeta^2+\chi^2}{1-\chi^2}}\,,
\end{equation}
\begin{equation}
\label{h_phi_obl}
h_\phi \equiv {h_\phi(\zeta,\chi)} =
\kappa_o \sqrt{(\zeta^2 + 1)(1-\chi^2)}\,,
\end{equation}
\end{subequations}
are the scale factors corresponding to the $\mathbf{\hat e}_\zeta$, 
$\mathbf{\hat e}_\chi$, and $\mathbf{\hat e}_\phi$ directions, 
respectively (Ref. \cite[(a)]{Abramowitz_book}). By using the 
orthogonality of the Legendre polynomials 
[Eq. (\ref{Legendre_ortho})], and by combining 
Eqs. (\ref{series_rho_oblate})-(\ref{h_zeta}), one finds the 
following expression for the coefficients $m_\ell$:
\begin{eqnarray}
\label{expr_m_ell}
m_\ell(\zeta_0,\chi_0) &=& 
-\dfrac{\nu_B \sigma R_1}{D} \,
\dfrac{(2 \ell + 1) \epsilon_\ell(\zeta_0,\chi_0)}
{2\, \zeta_0 \sqrt{\zeta_0^2 + 1}\,
[\partial_\zeta Q_\ell(\mathrm{i} \zeta)]_{\zeta = \zeta_0}}
\nonumber\\
&:=& - \dfrac{\nu_B \sigma R_1}{D} {\tilde m}_\ell(\zeta_0,\chi_0) \,,
\end{eqnarray}
where
\begin{eqnarray}
\label{epsilon_ell}
\epsilon_\ell(\zeta_0,\chi_0) &=&
\int\limits_{-1}^{1} d\chi\,
\sqrt{\zeta_0^2 + \chi^2} \,\Upsilon(\chi;\chi_0) P_\ell(\chi)
\nonumber\\
&=& \int\limits_{-1}^{\chi_0} d\chi\,
\sqrt{\zeta_0^2 + \chi^2} \,P_\ell(\chi)\,
\end{eqnarray}
so that the coefficients ${\tilde m}_\ell(\zeta_0,\chi_0)$ are 
dimensionless.

\subsection{Phoretic velocity of a prolate object.}
\label{phoretic_vel_prol_shape}
From differential geometry one has \cite{book_diff_geom}:\hfill\\
(i) $\nabla^\Sigma = \mathbf{\hat e}_\eta \,\dfrac{1}{h_\eta}
\dfrac{\partial}{\partial \eta} + 
\mathbf{\hat e}_\phi \,\dfrac{1}{h_\phi}
\dfrac{\partial}{ \partial \phi}$ and the surface area element is
$d\Sigma = h_\eta\,h_\phi\,d\eta\,d\phi$ because the plane tangent to
the surface of the particle is spanned by the unit vectors 
$\mathbf{\hat e}_\eta$ and $\mathbf{\hat e}_\phi$ of the $\eta$ and 
$\phi$ directions,\hfill\\
(ii) $\mathbf{\hat n} \cdot \mathbf{r} = 
\mathbf{\hat e}_\xi \cdot \mathbf{r} =
\dfrac{x \partial_\xi x + y \partial_\xi y + z \partial_\xi z}{h_\xi}
= \dfrac{\kappa_p^2 \xi}{h_\xi}$, and\\
(iii) $\mathbf{\hat e}_z \cdot \nabla^\Sigma \rho(\xi,\eta) =
(\mathbf{\hat e}_z \cdot \mathbf{\hat e}_\eta)\,
\dfrac{\partial_\eta \rho(\xi,\eta)}{h_\eta} =
\dfrac{\kappa_p \xi}{h_\eta^2}\,
\partial_\eta \rho(\xi,\eta)$\,.\hfill\\
By using the expression in Eq. (\ref{velocity}) for the phoretic 
velocity of a \textit{pr}olate object one obtains:
\begin{eqnarray}
\label{velocity_prolates}
V_{pr} &=& \dfrac{b \,\xi_0^2}{2 R_1}
\,\int\limits_{-1}^{1} d\eta\, \dfrac{1-\eta^2}{\xi_0^2 -\eta^2}
\,\partial_\eta \rho(\xi_0,\eta)\nonumber\\
&=& \dfrac{b \,\xi_0^2 \, (\xi_0^2 -1)^2}{R_1} \,
\,\int\limits_{-1}^{1} d\eta\, \dfrac{\eta}{(\xi_0^2 -\eta^2)^2} \,
\rho(\xi_0,\eta)\nonumber\\
&=& - 2 V_0 \,\xi_0^2 \, (\xi_0^2 -1)\,
\sum_{\ell \geq 0} \,[{\tilde c}_{2 \ell + 1}(\xi_0,\eta_0)\,
\nonumber\\
&\times&  Q_{2 \ell + 1}(\xi_0)
\,\int\limits_{0}^{1} d\eta\, \dfrac{\eta}{(\xi_0^2 -\eta^2)^2}
\,P_{2\ell+1} (\eta)\,]\,.
\end{eqnarray}
where
\begin{equation}
\label{def_V0}
V_0 = \dfrac{b \nu_B \sigma}{D}
\end{equation}
defines the velocity scale. $V_0$ is expected to be of the order of 
$\mu$m/s, but because of its dependence on $b$ it is difficult to 
provide a theoretical estimate for it. The second equality follows 
from an integration by parts, while the third one uses the series
expansion in Eq. (\ref{series_rho_prolate}) for $\rho(\xi_0,\eta)$,
Eqs. (\ref{parameter_k}) and (\ref{def_xi0}) to replace $\kappa_p$
and $s_r$, respectively, and the fact that $P_\ell(\eta)$ is an
even (odd) function of $\eta$ for $\ell$ even (odd). After
replacing ${\tilde c}_\ell$, $\xi_0$, and $\eta_0$ by the
corresponding expression in Eqs. (\ref{expr_c_ell}), (\ref{def_xi0}),
and (\ref{def_eta0}), one obtains as final result the phoretic 
velocity $V_{pr} \equiv V_{pr}(s_r,s_h;V_0)$ as a function of the 
geometrical parameters $s_r$ and $s_h$, as well as of the velocity 
scale $V_0$. 

\subsection{Phoretic velocity of an oblate object.}
\label{phoretic_vel_obl_shape}
Similar to the calculation in Subsec. \ref{phoretic_vel_prol_shape}, 
for an oblate shape one has \cite{book_diff_geom}:\hfill\\
(i) $\nabla^\Sigma = \mathbf{\hat e}_\chi \,\dfrac{1}{h_\chi}
\dfrac{\partial}{\partial \chi} + 
\mathbf{\hat e}_\phi \,\dfrac{1}{h_\phi}
\dfrac{\partial}{ \partial \phi}$ and the surface area element is
$d\Sigma = h_\chi\,h_\phi\,d\chi\,d\phi$,\hfill\\
(ii) $\mathbf{\hat n} \cdot \mathbf{r} = 
\mathbf{\hat e}_\zeta \cdot \mathbf{r} =
\dfrac{x \partial_\zeta x + y \partial_\zeta y + z \partial_\zeta z}
{h_\zeta}
= \dfrac{\kappa_o^2 \zeta}{h_\zeta}$, and\\
(iii) $\mathbf{\hat e}_z \cdot \nabla^\Sigma \rho(\zeta,\chi) =
(\mathbf{\hat e}_z \cdot \mathbf{\hat e}_\chi)\,
\dfrac{\partial_\chi \rho(\zeta,\chi)}{h_\chi} =
\dfrac{\kappa_o \zeta}{h_\chi^2}\,
\partial_\chi \rho(\zeta,\chi)$\,.\hfill\\
By using the expression in Eq. (\ref{velocity}) for the phoretic 
velocity of an \textit{ob}late object one obtains:
\begin{eqnarray}
\label{velocity_oblates}
V_{ob} &=& \dfrac{b \,\zeta_0^2}{2 R_1}
\,\int\limits_{-1}^{1} d\chi\, \dfrac{1-\chi^2}{\zeta_0^2 + \chi^2}
\,\partial_\chi \rho(\zeta_0,\chi)\nonumber\\
&=& \dfrac{b \,\zeta_0^2 \, (\zeta_0^2 +1)^2}{R_1} \,
\,\int\limits_{-1}^{1} d\chi\, \dfrac{\chi}{(\zeta_0^2 +\chi^2)^2} \,
\rho(\zeta_0,\chi)\nonumber\\
&=& - 2 V_0 \,\zeta_0^2 \, (\zeta_0^2 + 1)\,
\sum_{\ell \geq 0} \,[{\tilde m}_{2 \ell + 1}(\zeta_0,\chi_0)\,
\nonumber\\
&\times& Q_{2 \ell + 1}(\mathrm{i} \zeta_0)
\,\int\limits_{0}^{1} d\chi\, \dfrac{\chi}{(\zeta_0^2 +\chi^2)^2}
\,P_{2\ell+1} (\chi)\,]\,.
\end{eqnarray}
After replacing ${\tilde m}_\ell$, $\zeta_0$, and $\chi_0$ by the
corresponding expressions in Eqs. (\ref{expr_m_ell}), (\ref{def_zeta0}),
and (\ref{def_chi0}), one obtains as final result the phoretic 
velocity $V_{ob} \equiv V_{ob}(s_r,s_h;V_0)$. Similar to the case of a 
prolate object, the terms with even $\ell$ do not contribute to the 
phoretic velocity, i.e., $m_{\ell~\mathrm{even}} = 0$ because 
$P_\ell(\eta)$ is an even (odd) function of $\chi$ for $\ell$ even 
(odd). Note that the expression in Eq. (\ref{velocity_oblates}) can be 
obtained from Eq. (\ref{velocity_prolates}) by the mapping 
$\xi \mapsto \mathrm{i} \zeta$; this is in agreement with similar 
observations regarding solutions of the Laplace equation in 
spheroidal coordinates (see, e.g., Ref. \cite{prolates_to_oblates}), 
which is a welcome consistency check for our results.
\section{Discussion}
\label{discuss}

We first note that the length scale $R_1$ does not enter explicitly 
into the final expression for the velocity 
[Eqs. (\ref{velocity_prolates}) and (\ref{velocity_oblates})] which 
shows that within the assumptions of the model the phoretic velocity 
of objects with the same aspect ratio but different linear sizes is 
the same (in agreement with the conclusions of 
Ref. \cite{Golestanian_2007} for spheres and cylindrical rods).

The second observation concerns a symmetry with respect to the 
area covered by the catalyst. The velocity depends on the ratio 
$s_h$, or equivalently $\eta_0$ or $\chi_0$, via the integral in 
Eq. (\ref{gamma_ell}) and Eq. (\ref{epsilon_ell}), respectively, 
which enters into the coefficients of the corresponding series 
expansion of the density. With $P_\ell(-\eta)= -P_\ell(\eta)$ for 
$\ell$ odd Eq. (\ref{gamma_ell}) yields for $0 \leq w \leq 1$ 
and $\ell$ odd
\begin{eqnarray}
\label{sh_symmetry_gamma}
\gamma_\ell(\xi_0,\eta_0 =  - w) &=& \int\limits_{-1}^{-w} d\eta \,
\sqrt{\xi_0^2 - \eta^2} P_\ell(\eta) \nonumber\\
&=& 
\int\limits_{-1}^{w} d \eta \, \sqrt{\xi_0^2 - \eta^2}
P_\ell(\eta) \nonumber\\
&=& \gamma_\ell(\xi_0,\eta_0 = w)\,,
\end{eqnarray}
because $\int\limits_{-1}^{1} d\eta \, \sqrt{\xi_0^2 - \eta^2} 
P_\ell(\eta) = 0$  for $\ell$ odd. A similar relation holds for the 
case of an oblate shape: 
\begin{equation}
\label{sh_symmetry_epsilon}
\epsilon_\ell(\zeta_0,\chi_0 =  - w) = 
\epsilon_\ell(\zeta_0, \chi_0 = w)\,.
\end{equation}
Since $s_h = 1 + \eta_0 \equiv 1 + \chi_0$, this means that the 
coefficients in the series expansion in 
Eqs. (\ref{velocity_prolates}) and (\ref{velocity_oblates}), and 
thus the velocity of the particle, are the same if the catalytic 
coverage is less or more than half of the particle by the same amount 
$w$. Equivalently, it means that if the inert and the catalytic 
characteristics of the two parts of the surface are interchanged, 
the velocity is the same but the direction of motion is reversed. 
Therefore, as a function of $s_h$, $V_{pr}$ and $V_{ob}$ 
have an extremum at $s_h = 1~(\eta_0 = \chi_0 = 0)$, i.e., if half 
of the particle surface is covered by the catalyst. For $s_h = 0$ 
(no catalyst) or $s_h = 2$ (entire particle surface covered by the 
catalyst) and a finite number density $\sigma$ of catalytic sites, 
the coefficients $c_{2 \ell +1}$ and $m_{2 \ell +1}$ are identically 
zero and the velocities predicted by Eqs. (\ref{velocity_prolates}) 
and (\ref{velocity_oblates}) vanish as expected because in these two 
cases there are no gradients of the density of the product molecules 
along the surface of the particle.

The dependences on $\xi_0$ and $\eta_0$ as well as on $\zeta_0$ 
and $\chi_0$ (i.e., on $s_r$ and $s_h$) of the series representations 
of the velocity [Eqs. (\ref{velocity_prolates}) and 
(\ref{velocity_oblates})] are very complicated and in the general 
case we have not been able to obtain a solution in closed form. 
Therefore, we shall study numerically the cases of generic prolate 
($0 < s_r \leq 1$) and oblate ($1 < s_r \leq \infty$) shapes, and we 
shall complement the analysis with analytical results for the 
limiting cases of a spherical ($s_r = 1$), a needle-like 
($s_r \to 0$), and a disk-like ($s_r \to \infty$) shaped particle.

\begin{figure}[!htb]
\begin{center}
\includegraphics[width=.8\linewidth]{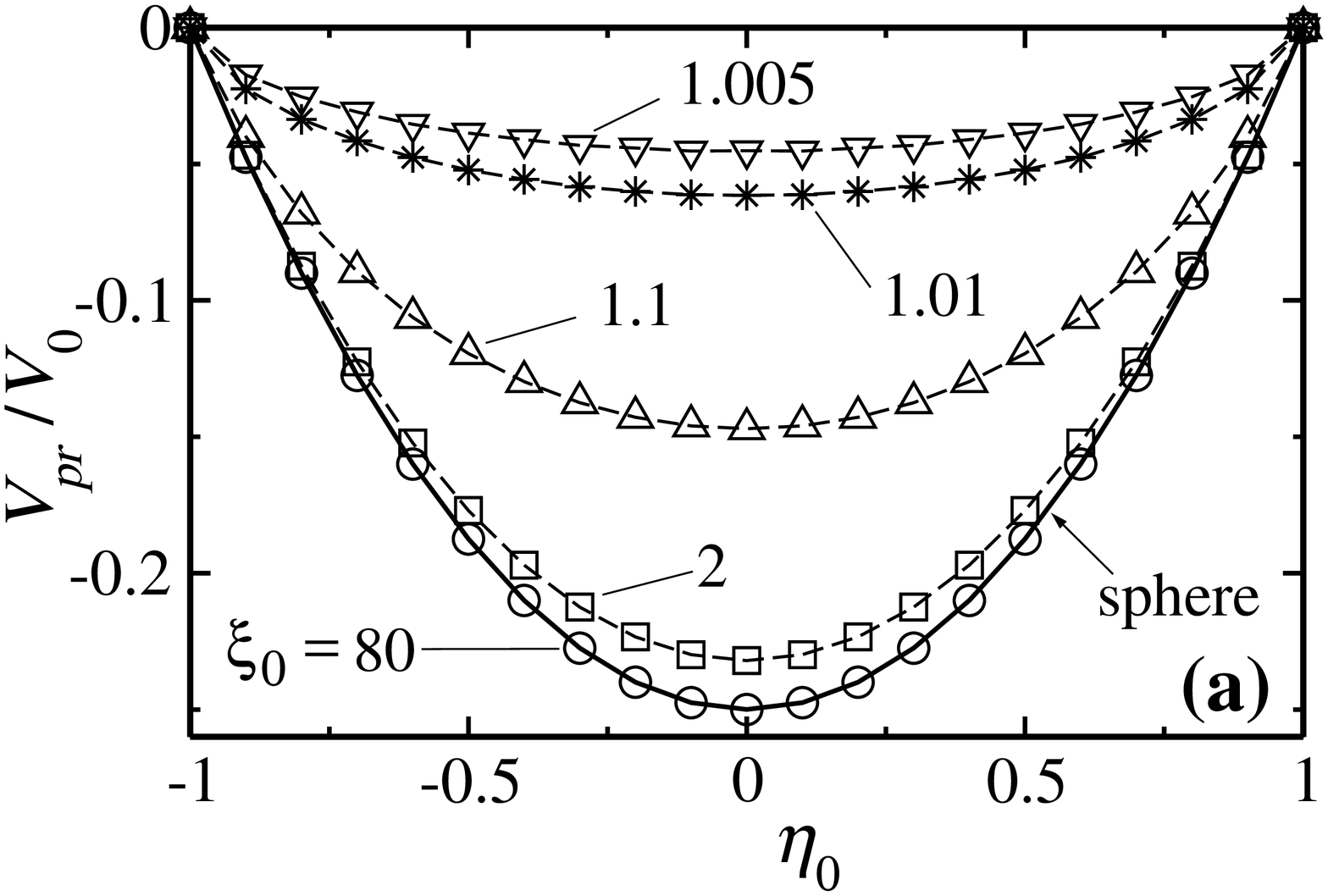}\hfill\\
\includegraphics[width=.8\linewidth]{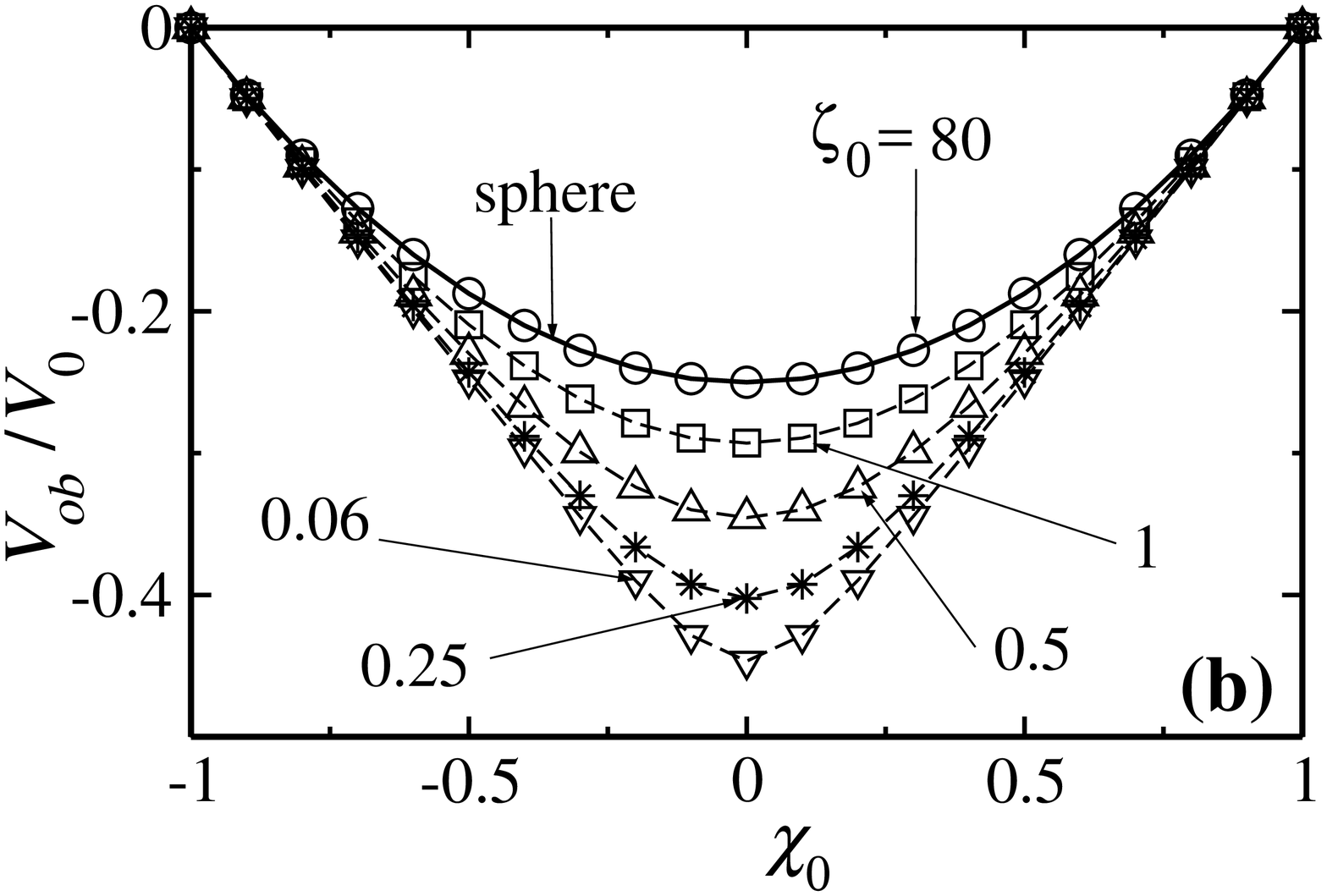}\hfill
\end{center}
\caption
{
\label{fig4}
Dependence of the scaled phoretic velocity 
[Eqs. (\ref{velocity_prolates}),  (\ref{velocity_oblates}), and 
(\ref{def_V0})] on the fraction $s_h = 1 + \eta_0 \equiv 1 + \chi_0$ 
of the surface of the particle covered by the catalyst 
($\eta_0,~\chi_0 = -1, 0, 1$ correspond to a particle surface with 
no catalyst, with the lower half covered by the catalyst, and 
completely covered by the catalyst, respectively), for (a) a 
prolate-shaped particle with aspect ratio parameter $\xi_0 = 
(1-s_r^2)^{-1/2}$ = 80  ($s_r = 0.9999$), 2 ($s_r = 0.866$), 
1.1 ($s_r = 0.42$), 1.01 ($s_r = 0.14$), and 1.005 ($s_r = 0.099$); 
and (b) an oblate-shaped particle with aspect ratio parameter 
$\zeta_0 =  (-1+s_r^2)^{-1/2}$ = 80 ($s_r = 1.00008$), 1.0 
($s_r = 1.41$), 0.5 ($s_r = 2.24$), 0.25 ($s_r = 4.12$), and 0.06 
($s_r = 16.7$). The numerical results are shown as symbols; the 
dashed lines connecting the symbols are guides for the eye. Also 
shown (solid line) is the exact result [Eq. (\ref{velocity_sphere})] 
corresponding to a spherical particle ($\xi_0,~\zeta_0 \to \infty$).
}
\end{figure}

For given $\xi_0$ and $\eta_0$ ($\zeta_0$ and $\chi_0$) we approximate 
the velocity $V_{pr}$ ($V_{ob}$) by keeping terms up to $\ell = 15$ 
in Eq. (\ref{velocity_prolates}) [Eq. (\ref{velocity_oblates})]. 
This provides a good approximation for all values 
(a) $1.001 \leq \xi_0 \leq 80$ (i.e., $0.0451 \leq s_r \leq 0.9999$) 
and $-1 \leq \eta_0 \leq 1$ (prolate shapes) and (b)
$0.01 \leq \zeta_0 \leq 80$ (i.e., $1.00008 \leq s_r \leq 100$) and 
$-1 \leq \chi_0 \leq 1$ (oblate shapes) which we have tested (in the 
sense that including five additional terms leads to changes in the 
value of the velocity smaller than $10^{-6}$) \cite{note_3}. The 
results for $V_{pr}/V_0$ and $V_{ob}/V_0$ are shown in 
Fig. \ref{fig4} as functions of $\eta_0$ and $\chi_0$ for several 
values of $\xi_0$ and $\zeta_0$, respectively. The data span the 
whole range of interest, from $s_r \ll 1$ (i.e., $\xi_0 \gtrsim 1$), 
corresponding to a ``cigar'' shaped rod, to $s_r \gg 1$ (i.e., 
$\zeta_0 \ll 1$), corresponding to a disk, through $s_r \simeq 1$ 
(i.e., $\xi_0 \gg 1$ or $\zeta_0 \gg 1$) corresponding to a slightly 
deformed sphere. 

%%%%%%%%%%%%%%%%%%%%%%%%%
%%%%%%%%%%%%%%%%%%%%%%%%%
Figure \ref{fig4} shows that the dependence of $V/V_0$ on $\eta_0$ 
(or $\chi_0$) is qualitatively the same for all values of $\xi_0$ 
and $\zeta_0$. The ratio $V/V_0$ is always negative, which means 
that $V$ and $V_0$ have opposite signs. Therefore, for repulsive 
effective interactions between the particle and the product 
molecules, i.e., in the case that $b < 0$ and thus $V_0 < 0$, $V$ 
is positive and the particle moves in the positive $z$ direction, 
i.e., ``away'' from the catalyst covered end. Similarly, for 
attractive effective interactions the motion will be directed 
``towards'' the catalyst covered end. As expected the velocity 
curves are symmetric with respect to $\eta_0 = 0$ and $\chi_0 = 0$ 
(i.e., $s_h = 1$), which is a highly welcome check for the 
numerical calculations. For a prolate, the minimum at $\eta_0 = 0$ 
slowly rises towards zero upon decreasing $\xi_0$ towards 
$\xi_0 = 1^+$, so that the velocity vanishes for 
$s_r = \sqrt{1-\xi_0^{-2}} \to 0$; as discussed below, this is in 
qualitative agreement with 
Refs. \cite{Paxton_2004,Saidulu_2008,Golestanian_2007}. In contrast, 
for an oblate the absolute value of the velocity increases upon 
decreasing $\zeta_0$ (at fixed $\chi_0$) towards a non-zero and 
finite limiting value which corresponds to that for an infinitely 
thin disk (oblate of equatorial diameter much larger than the 
polar one) [see, c.f., Subsec. \ref{equiv_with_disk}]. 
(Note that the velocity of such an object remains finite because 
the product molecules can still ``escape'' to both sides of the 
object due to the assumption of an unbounded solvent.) Therefore, 
we conclude that for a given coverage by catalyst a flatter shape 
of the particle leads to an increased phoretic velocity. This can 
be understood intuitively by analyzing the case of a single, 
point-like catalyst source placed at one of the poles of the 
particle. The spherically symmetric product number density 
distribution due to a point-like source but in the absence of the 
particle will be deformed over a larger extent by an impenetrable 
boundary which is more elongated transversal to the radial direction 
than by one more elongated along the radial direction, and thus the 
gradient along the surface of an oblate particle is expected to be 
larger than the one corresponding to a prolate particle. 
Finally, we note that (i) for large values of $\xi_0$ or $\zeta_0$ 
the numerical solutions are in perfect agreement with the known 
analytical result [Eq. (\ref{velocity_sphere})] for a sphere 
\cite{Golestanian_2007}; and that (ii) for very elongated objects 
(rods or oblates) the results should be considered only as a 
first-order approximation. This is the case because the analysis 
leading to the expression in Eq. (\ref{slip_vel}) approximates the 
flow within the thin surface film as locally planar; such an 
assumption may break down near the poles for prolates or near the 
equator for oblates. There for very elongated particles ($s_r \ll 1$ 
or $s_r \gg 1$) the curvature of the surface of the particle is 
very high and cannot be neglected \textit{a priori}.

Further insight can be obtained from an analytical study of the 
behavior in the limiting cases of a slightly deformed sphere: 
$\xi_0 \gg 1$ ($s_r \lesssim 1$) or $\zeta_0 \gg 1$ 
($s_r \gtrsim 1$), of a needle-like shape:  
$\xi_0 \gtrsim 1$ ($s_r \ll 1$), and of 
a disk: $\zeta_0 \ll 1$ ($s_r \gg 1$) which we shall discuss now. 

\subsection{Limit of a slightly deformed sphere}
\label{equiv_with_sphere}

We first consider the case in which the limit towards a sphere is 
taken through prolate shapes, i.e., $s_r \to 1^-$. Noting that in 
the limit of a sphere $s_r = R_2/R_1 \to 1^-$, which is equivalent 
to $\xi_0 \to \infty$ [see Eq. (\ref{def_xi0})], the integral in the 
last line of  Eq. (\ref{velocity_prolates}) can be approximated as:
\begin{eqnarray}
&&\int\limits_{0}^{1} d\eta\, \dfrac{\eta}{(\xi_0^2 -\eta^2)^2}
\,P_{2\ell+1} (\eta) \nonumber\\
&&\hspace*{1.cm}
= \xi_0^{-4} 
\left[\int\limits_{0}^{1} d\eta\, \eta P_{2\ell+1} (\eta) + 
{\cal O}(\xi_0^{-2}) \right]\nonumber\\
&&\hspace*{1.cm} 
= \xi_0^{-4} \,\left[\dfrac{1}{3} \,\delta_{\ell,0} + 
{\cal O}(\xi_0^{-2})\right]\,.\nonumber
\end{eqnarray}
Accordingly, Eq. (\ref{velocity_prolates}) reduces to
\begin{equation}
\label{V_prol_sphere_firstep}
\dfrac{V_{pr}(\xi_0 \gg 1)}{V_0} 
\simeq - \left[\xi_0^{-2} (1-\xi_0^{-2})^{1/2}
\dfrac{\gamma_1(\xi_0) Q_1(\xi_0)}{Q'_1(\xi_0)}
\right]_{\xi_0 \gg 1} \,.
\end{equation}
Furthermore, Eq. (\ref{gamma_ell}) implies
\begin{eqnarray}
&&\gamma_1(\xi_0 \gg 1) = \xi_0\,
\left[
\int\limits_{-1}^{1} d\eta\, \Upsilon(\eta;\eta_0) P_1(\eta)
\right.\nonumber\\
&& \left. + \dfrac{1}{2}\,\xi_0^{-2} 
\int\limits_{-1}^{1} d\eta\,\eta^2\, \Upsilon(\eta;\eta_0) P_1(\eta) 
+ {\cal O}(\xi_0^{-4}) \right];
\end{eqnarray}
note that the first integral on the right hand side is equal to 
twice the velocity (in units of the characteristic velocity $V_0$) 
corresponding to a spherical particle 
[Eq. (\ref{velocity_sphere}), Appendix \ref{app_B})]. Due to
\begin{equation}
\left[\dfrac{Q_1(\xi_0)}{\xi_0 Q'_1(\xi_0)}
\right]_{\xi_0 \gg 1} = -\dfrac{1}{2} + \dfrac{3}{10}\,\xi_0^{-2} + 
{\cal O}(\xi_0^{-4})
\end{equation}
and $V_{sph}/V_0 = (\eta_0^2-1)/4$ one finds with $\eta_0^2 \leq 1$ 
[Eq.~(\ref{def_eta0})]
\begin{eqnarray}
\label{V_prol_sphere}
&&\dfrac{V_{pr}(\xi_0 \gg 1)}{V_0} \simeq 
\left(1- \dfrac{1}{2}\,\xi_0^{-2}\right)\,\nonumber\\
&&\times\,
\left[\dfrac{V_{sph}}{V_0} + \xi_0^{-2} \, 
\int\limits_{-1}^{1} d\eta\, 
\dfrac{5 \eta^2 - 6}{20} \, \Upsilon(\eta;\eta_0) P_1(\eta)
\right]+{\cal O}(\xi_0^{-4})
\nonumber\\
&&= \dfrac{V_{sph}}{V_0}\,
\left(1 - \dfrac{1}{2}\,\xi_0^{-2}\right)\, 
\left(1 + \dfrac{5 \eta_0^2 - 7}{20}\,\xi_0^{-2}\right)
+ {\cal O}(\xi_0^{-4})\nonumber\\
&&\simeq \dfrac{V_{sph}}{V_0}\,
\left(1 - \dfrac{17-5 \eta_0^2}{20}\,\,\xi_0^{-2}\right)
\geq \dfrac{V_{sph}}{V_0}\,,
\end{eqnarray}
where the last inequality follows because the velocity ratios are 
negative quantities. Equation (\ref{V_prol_sphere}) thus implies that 
the absolute value of the velocity of a sphere slightly deformed 
towards a prolate shape ($\xi_0 \gg 1$) is \textit{smaller} than 
that of a sphere and becomes equal to it in the limit of a 
vanishing deformation ($\xi_0 \to \infty$), in agreement with the 
numerical findings shown in Fig. \ref{fig4}(a).

A similar calculation can be performed for the case in which the 
limit is taken through oblate shapes, i.e., $s_r \to 1^+$, which is 
equivalent to $\zeta_0 \to \infty$ [see Eq. (\ref{def_zeta0})], by 
starting from the integral in the last line of 
Eq. (\ref{velocity_oblates}). Following the same steps as above and  
noting that 
\begin{equation}
\left[
\dfrac{Q_1(\mathrm{i} \zeta_0)}{\zeta_0 \,
[\partial_\zeta Q_1(\mathrm{i} \zeta)]_{\zeta_0}} 
\right]_{\zeta_0 \gg 1} 
= -\dfrac{1}{2} + \dfrac{3}{10}\,\zeta_0^{-2} + 
{\cal O}(\zeta_0^{-4})
\end{equation}
we obtain with $\chi_0^2 \leq 1$ due to $\chi_0 = \eta_0$ [see 
Eqs. (\ref{def_eta0}) and (\ref{def_chi0})]
\begin{eqnarray}
\label{V_obl_sphere}
&&\dfrac{V_{ob}(\zeta_0 \gg 1)}{V_0} \simeq
\left(1 + \dfrac{1}{2}\,\zeta_0^{-2}\right)\,\nonumber\\
&&\times\,
\left[\dfrac{V_{sph}}{V_0} - \zeta_0^{-2} \, 
\int\limits_{-1}^{1} d\chi\, \dfrac{5 \chi^2 - 6}{20} \, 
\Upsilon(\chi;\chi_0) P_1(\chi)\right]+{\cal O}(\zeta_0^{-4})
\nonumber\\
&&= \dfrac{V_{sph}}{V_0}\,
\left(1 + \dfrac{1}{2}\,\zeta_0^{-2}\right)\, 
\left(1 - \dfrac{5 \chi_0^2 - 7}{20}\,\zeta_0^{-2}\right)
+ {\cal O}(\zeta_0^{-4})\nonumber\\
&&\simeq \dfrac{V_{sph}}{V_0}
\left(1 + 
\dfrac{17-5 \chi_0^2}{20}\,
\zeta_0^{-2} \right)
\leq \dfrac{V_{sph}}{V_0}\,,
\end{eqnarray}
because the velocity ratios are negative quantities. Thus the 
absolute value of the velocity of a sphere slightly deformed 
towards an oblate shape 
($\zeta_0 \gg 1$) is \textit{larger} than that of a sphere, while 
in the limit of a vanishing deformation ($\xi_0 \to \infty$) it 
indeed reduces to the velocity of a spherical particle, again in 
agreement with the numerical findings shown in Fig. \ref{fig4}(b).

\subsection{Limit of a needle-like particle}
\label{equiv_with_rod}

In this subsection we focus on estimating the asymptotic behavior 
of the velocity of a half-covered ($s_h = 1,~\eta_0 = 0$) prolate 
particle with increasing elongation towards a needle-like shape 
($s_r \ll 1$, i.e., $\xi_0 \gtrsim 1$). As shown in 
Fig. \ref{fig4}(a), for given $s_r \leq 1$ 
(or, equivalently, $\xi_0 > 1$) the absolute value of the velocity 
for any value $\eta_0 \neq 0$ is smaller than the one for 
$\eta_0 = 0$. Thus the asymptotic behavior of 
$|V_{pr}(\xi_0 \to 1^+, \eta_0)|$ has an upper bound given by 
$|V_{pr}(\xi_0 \to 1^+, \eta_0 = 0)|$.

Eq. (\ref{velocity_prolates}) may be re-written as:
\begin{eqnarray}
\label{V_prol_rod_firstep}
& & V_{pr}(\xi_0,\eta_0 = 0)
=
\dfrac{b \xi_0^2}{2 R_1} 
\int\limits_{-1}^1 d \eta \,
\left(1 - \dfrac{\xi_0^2-1}{\xi_0^2-\eta^2}\right) 
\partial_\eta \rho(\xi_0,\eta)\nonumber\\ 
&&\stackrel{Eq.\,(\ref{series_rho_prolate})}{=}
\dfrac{b \xi_0^2}{2 R_1}\,[\rho(\xi_0,\eta=1)-\rho(\xi_0,\eta=-1)]
\nonumber\\
&&~~~~~-\dfrac{b \xi_0^2}{2 R_1}\,\left(\xi_0^2-1\right)\nonumber\\
&&~~~~~~\times
\sum_{\ell~\textrm{odd}} c_\ell (\xi_0,\eta_0=0) 
Q_\ell(\xi_0)
\int\limits_{-1}^1 d \eta\,
\dfrac{1}{\xi_0^2-\eta^2} \dfrac{dP_\ell}{d\eta}\nonumber\\
&&\stackrel{Eq.\,(\ref{series_rho_prolate})}{=}
~\dfrac{b \xi_0^2}{R_1}
\sum_{\ell~\textrm{odd}} c_\ell (\xi_0,\eta_0=0) 
Q_\ell(\xi_0) \nonumber\\
&& ~~~~~~~\times \left[1-\dfrac{\xi_0^2-1}{2}
\int\limits_{-1}^1 d \eta\,
\dfrac{1}{\xi_0^2-\eta^2} \dfrac{dP_\ell}{d\eta}\right]\,,
\end{eqnarray}
where we have used $P_\ell(1) = 1$ \cite[(c)]{Abramowitz_book} and 
the fact that for even indices $\ell$ the integrals are identically 
zero because in this case $P_\ell(\eta)$ is an even function of 
$\eta$ so that the integrands are odd functions of $\eta$. For odd 
$\ell$ the derivative $dP_\ell/d\eta$ is a 
polynomial of order $\ell - 1$ containing only even powers of 
$\eta$. Thus it can be written as
\begin{equation}
\label{quotient}
\tilde p_{\ell-1}(\eta) := \dfrac{d\,P_\ell}{d\,\eta} = 
\tilde \beta_\ell(\xi_0) + (\xi_0^2-\eta^2) \tilde q_{\ell-3}(\eta;\xi_0) \,,
\end{equation}
where 
\begin{equation}
\label{beta_expresion}
\tilde \beta_{\ell} (\xi_0)= 
\left(\dfrac{d\,P_\ell}{d\,\eta}\right)_{\eta = \xi_0} \,.
\end{equation}
By construction the second term in Eq. (\ref{quotient}) must vanish 
for $\eta = \xi_0$. Since $dP_\ell/d\eta$ depends on $\eta$ only via 
$\eta^2$, this vanishing must exhibit a prefactor $\xi_0^2-\eta^2$ 
multiplying a polynomial $\tilde q_{\ell-3}(\eta;\xi_0)$ of degree 
$(\ell-3)$  with $\tilde q_{m < 0} \equiv 0$. 
Accordingly, the integral in the last equation of 
Eq. (\ref{V_prol_rod_firstep}) is approximated by
\begin{eqnarray}
\label{Jl_approx}
\int\limits_{-1}^1 d \eta\,
\dfrac{1}{\xi_0^2-\eta^2} \dfrac{dP_\ell}{d\eta} && =- 
\dfrac{\tilde \beta_{\ell}(\xi_0)}{\xi_0} \ln\dfrac{\xi_0-1}{\xi_0+1}
+ \int\limits_{-1}^1 d \eta\, \tilde q_{\ell-3}(\eta;\xi_0) \nonumber\\
&& \stackrel{\xi_0 \to 1^+}{\to} -\tilde \beta_{\ell}(\xi_0 = 1) 
\ln(\xi_0-1)\,.
\end{eqnarray}
This implies that in the limit $\xi_0 \to 1^+$ the second term in 
the square bracket in Eq. (\ref{V_prol_rod_firstep}) vanishes 
$\sim (\xi_0-1) \ln(\xi_0-1)$ and therefore it is a subdominant 
contribution to the quantity in the square bracket. Thus 
Eqs. (\ref{V_prol_rod_firstep}) and (\ref{Jl_approx}) lead to 
the following approximation for the velocity of a very elongated 
prolate towards a needle-like shape ($\xi_0 \to 1^+$):
\begin{eqnarray}
\label{V_prol_rod_first_approx}
&& V_{pr}(\xi_0 \gtrsim 1,\eta_0 = 0) \to
\dfrac{b \xi_0^2}{R_1} \nonumber\\
&& \times \sum_{\ell~\textrm{odd}} c_\ell (\xi_0,\eta_0=0) Q_\ell(\xi_0) 
\left[1 + {\tilde \beta}_\ell (1)\,
(\xi_0-1)\ln(\xi_0-1)\right]\nonumber\\
&& \to \dfrac{b \xi_0^2}{R_1} 
\sum_{\ell~\textrm{odd}} c_\ell (\xi_0,\eta_0=0) Q_\ell(\xi_0)
\nonumber\\ 
&& \stackrel{Eq.\,(\ref{expr_c_ell})}{\to} 
-\dfrac{V_0}{2\,\sqrt{2}\sqrt{\xi_0-1}}\nonumber\\
&& \times \sum_{\ell~\textrm{odd}} (2 \ell +1)
\left[
\dfrac{Q_\ell(\xi_0)}{Q'_\ell(\xi_0)} \,\gamma_\ell(\xi_0,\eta_0=0) 
\right]_{\xi_0 \gtrsim 1}\,.
\end{eqnarray}
Within this approximation the velocity of a very thin, needle-like 
prolate object is proportional to the difference between the values 
of the product molecules density at the two ends ($\eta = \pm 1$) of 
the object. (Note that because of this subtraction only those terms 
with odd $\ell$ occur in the series representation of the velocity.) 
This is similar to the results for a cylindrical thin 
rod postulated in Ref. \cite{Paxton_2004} and derived in 
Ref. \cite{Golestanian_2007} by invoking a ``slender-body'' 
approximation.

Although the above approximation leads to a significantly 
simplified expression for the velocity, we have been unable to 
further simplify the resulting series and thus we have studied it 
numerically. The conclusion of this analysis, the details of which 
are presented in Appendix \ref{app_C}, is that for 
$\xi_0 \gtrsim 1$ the series in Eq. (\ref{V_prol_rod_first_approx}) 
behaves as
\begin{eqnarray}
\label{f_tilde}
{\tilde f}(\xi_0) &:=& \sum_{\ell~\textrm{odd}} (2 \ell +1)
\left[\dfrac{Q_\ell(\xi_0)}{Q'_\ell(\xi_0)} 
\gamma_\ell(\xi_0,\eta_0 = 0)\right]_{\xi_0 \gtrsim 1} \nonumber\\
&\to& (\xi_0-1) \ln(\xi_0-1) f(\xi_0) \,,
\end{eqnarray}
where 
\begin{equation}
\label{f_expression}
f(\xi_0 \gtrsim 1) \simeq \, -1.5 \times [-\ln(\xi_0-1)]^{-0.9}\,.
\end{equation}
Eqsuations (\ref{V_prol_rod_first_approx}) and (\ref{f_expression}) 
therefore render in the limit $\xi_0 \to 1^+$ 
\begin{eqnarray}
\label{V_prolate_rod}
&& \dfrac{V_{pr}(\xi_0 \gtrsim 1,\eta_0 = 0)}{V_0} 
\simeq \dfrac{1.5}{2 \sqrt{2}} \,\,
\sqrt{\xi_0-1} \,\,\ln (\xi_0-1) \nonumber\\
&& \times \,[-\ln(\xi_0-1)]^{-0.9} < 0 \,.
\end{eqnarray}
In agreement with the behavior observed for a general prolate shape 
[see Fig. \ref{fig4}(a)], the velocity (in units of $V_0$) is 
negative. For $\xi_0 \to 1^+$, the velocity vanishes \textit{faster} 
than $\sqrt{\xi_0-1} \,\ln (\xi_0-1)$, i.e., more rapidly than the 
behavior predicted by Refs. \cite{Golestanian_2007,Paxton_2004}. This 
difference can be either due to an intrinsic difference 
between the cylinder-like and the needle-like shapes, or, most likely, 
due to the additional approximations employed in Refs. 
\cite{Golestanian_2007,Paxton_2004} upon computing the solute density 
distribution (such as using a distribution of point sources in 
unbounded space and the absence of sources on the flat ends of the 
cylinder).

We note that in terms of an effective power law the value 0.9 of the exponent 
in Eq.~(\ref{f_expression}) provides a very good 
approximation for the behavior of $f(\xi_0)$ 
over the physically accessible range of values $\xi_0 \to 1^+$, which 
can be estimated to be bounded from below by $\xi_0 \simeq 1 + 10^{-9}$ 
[i.e., the value corresponding to a $10~\mu$m long carbon nanotube 
of 1 nm diameter ($s_r = 10^{-4}$)]. 
However, in a strictly mathematical sense, the limiting behavior of
$f(\xi_0 \to 1^+)$ appears to be not given by Eq.~(\ref{f_expression}) because one finds that the 
exponent decreases as 
the range $\xi_0$ under consideration corresponds to smaller and 
smaller values; e.g., the expoenent reaches the value 0.7 for 
$\xi_0 \gtrsim 1+ 10^{-128}$.

\subsection{Limit of a disk}
\label{equiv_with_disk}

In this subsection we focus on the case of a half-covered 
($s_h = 1,~\chi_0 = 0$), very flat oblate (disk-like shape, 
$\zeta_0 \ll 1$) because among the class of spheroids we have 
studied this is the one which exhibits the largest absolute value 
of the velocity.

For $\zeta_0 \ll 1$, the coefficients $\epsilon_n$ with an odd index 
$n = 2\ell+1$ [Eq. (\ref{epsilon_ell})] can be approximated by
\begin{eqnarray}
\label{eps_for_disk}
&& \epsilon_{2\ell + 1}(\zeta_0 \ll 1, \chi_0 = 0) 
\simeq
\int\limits_{-1}^{\chi_0 = 0}\,d\chi\, (-\chi) \,P_{2\ell + 1}(\chi) 
\nonumber\\
&& \hspace*{1cm}= -\dfrac{1}{2} \int\limits_{-1}^{1}\,d\chi\, P_1(\chi)\, 
P_{2 \ell+1}(\chi) = -\dfrac{1}{3}\,\delta_{\ell,0}\,.
\end{eqnarray}
(According to Subsec. \ref{phoretic_vel_obl_shape} the coefficients 
$\epsilon_{2 \ell}$ do not contribute to $V_{ob}$.) Therefore in 
this limiting case only the first term in the series representation 
given in Eq. (\ref{velocity_oblates}) contributes. Since
\begin{eqnarray}
\left.\dfrac{Q_1(\mathrm{i} \zeta_0)}
{[\partial_\zeta Q_1(\mathrm{i} \zeta)]_{\zeta = \zeta_0}}
\right|_{\zeta_0 \ll 1} &=& 
\left.\dfrac{-1 + \zeta_0 \mathrm{arccot}(\zeta_0)}
{-\dfrac{\zeta_0}{1+\zeta_0^2}+ 
\mathrm{arccot}(\zeta_0)}\right|_{\zeta_0 \ll 1} \nonumber\\
&\simeq& -\dfrac{2}{\pi}
\end{eqnarray}
and
\begin{equation}
\left.\int\limits_{0}^{1} 
\,d\chi\,\dfrac{\chi}{(\zeta_0^2+\chi^2)^2}\,P_1(\chi) 
\right|_{\zeta_0 \ll 1} 
\simeq \,\dfrac{1}{2}\, \zeta_0^{-1} \mathrm{arccot}(\zeta_0)\,,
\end{equation}
one obtains
\begin{eqnarray}
V_{ob}(\zeta_0 \ll 1, \chi_0 = 1)
&\simeq& -\dfrac{1}{2} \, \sqrt{1+\zeta_0^2}\, \,V_0 \nonumber\\
&\to& 
-\dfrac{1}{2} \, V_0~\mathrm{for}~\zeta_0 \to 0\,.
\end{eqnarray}
Thus in our context the maximal velocity of a spheroidal object is 
$|V_0/2|$ and it is realized for a particle with a disk-like shape 
and such that one side is covered by catalyst and the other one is 
inert. This is in agreement with the numerical results in 
Fig. \ref{fig4}.

The general case $\chi_0 \neq 0$ can be studied in a similar way, 
but due to the fact that in this case \textit{all} coefficients 
$\epsilon_\ell$ with $\ell$ odd  will contribute [see 
Eq. (\ref{epsilon_ell}) with $\chi_0 \neq 0$] to the series 
representation in Eq. (\ref{velocity_oblates}), the resulting 
velocity has to be calculated numerically.

\section{Summary}
\label{summary}

We have studied the diffusio-phoretic velocity of a 
spheroidal-shaped particle (Fig. \ref{fig1}) which self-propels by 
creating gradients of product molecules in a surrounding, unbounded 
solvent (Fig. \ref{fig3}). Our calculations yield the dependence of 
the velocity on 
the shape of the particle, i.e., the aspect ratio between the polar 
and the equatorial diameters, and on the fraction of the surface of 
the dissolved particle which is catalytically active in providing 
product molecules via chemical reactions.

The analysis of our model is based on recasting it in the framework
of the standard theory of phoresis. In this context we have 
critically analyzed the assumptions involved in such an approach, 
some of them already present in the standard theory, others arising 
as a result of the mapping of such ``active'' surface particles into 
the framework of a theory developed to describe the case of inert 
particles immersed in a pre-defined, externally controlled 
concentration gradient. Within the confines of the standard theory 
of phoresis, we have shown that, irrespective of the shape and of 
the fraction of the surface covered by the catalyst, the phoretic 
velocity of the particle depends on its geometry only via the aspect 
ratio but it is independent of its absolute size. The numerical 
analysis of the series representation for the phoretic velocity has 
been complemented by analytical results for the asymptotic cases of 
spherical and needle-like particles (Fig. \ref{fig2}). For a given 
coverage of the particle surface by the catalyst, the absolute value 
of the velocity of a prolate particle is maximal for an almost 
spherical particle and decreases gradually towards zero with 
increasing elongation of the prolate towards a needle-like shape 
[Fig. \ref{fig4}(a)]. We performed thorough numerical studies 
(Fig. \ref{fig5}) of the decay of the velocity with increasing 
elongation ($R_1 \gg R_2$, see Fig. \ref{fig1}) and concluded that 
over the whole physically accessible range 
$\xi_0 \gtrsim 1+ 10^{-9}$ it decays effectively 
$\sim \left(\dfrac{R_2}{R_1} \,\,\ln \, \dfrac{R_2}{R_1} \right)\,\,
\left(-\ln \, \dfrac{R_2}{R_1}\right)^{-0.9} = 
-\dfrac{R_2}{R_1} \,\,\left(-\ln \, \dfrac{R_2}{R_1}\right)^{0.1}
$, 
which is faster than the previously 
predicted $\left(\dfrac{R_2}{R_1} \ln \, \dfrac{R_2}{R_1}\right)$ 
behavior \cite{{Golestanian_2007,Paxton_2004,Saidulu_2008}}; this 
difference is most likely due to the approximations employed in 
these previous calculations. Our numerical study also indicates that 
this conclusion of a faster decay than previously predicted holds in 
the limit (of pure mathematical interest) $\xi_0 \to 1^+$ (see the 
inset in Fig. \ref{fig5}), but we could not determine the exact 
analytical form of this decay. In contrast, 
an oblate-shaped particle moves faster with increasing flattening 
from an almost spherical towards a disk-like shape 
[Fig. \ref{fig4}(b)]. For a given shape the maximal absolute value of 
the velocity $V$ in units of the characteristic velocity $V_0$ is 
always attained at half-coverage by the catalyst and varies from 
$0$ (needle) over $1/4$ (sphere) to $1/2$ (disk). Therefore, 
experimental realizations of such self-propelled particles call for 
a compromise on one hand between the increased speed of flatter 
particles and their decreased uni-directionality due to, e.g., the 
unavoidable thermal noise of the solvent and the density 
fluctuations of the product molecules, and on the other hand between 
the increased stability against rotations of more elongated 
particles and their reduced velocity. The results for the 
phoretic velocity $V/V_0$ shown in Fig. \ref{fig4} allow one to 
directly compare them with experimental realizations, e.g., confocal 
microscopy studies of micron sized needle-, sphere-, and disk-shaped 
particles, both qualitatively -- the symmetry with respect to the 
half-covered case and the dependence on the aspect ratio 
$s_r = R_2/R_1$ of the shape -- as well as quantitatively: the 
velocity of a half-covered disk is twice that of a spherical particle 
of the same radius.

As we pointed out in our discussion of the connection between the 
phoretic slip and the number density gradients of product molecules 
(Subsec. \ref{phor_slip_and_vel_subsec}), a natural extension of the 
present study would be to consider in detail the generic case in 
which the effective interaction potential between the product 
molecules and the particle differs on the catalyst-covered part of 
the particle surface from that on the bare one. Other questions of 
further interest are the influence of curvature on the phoretic 
velocity in the case of very elongated or very flat spheroidal 
particles and the influence of external boundaries on both the 
velocity and the uni-directionality of the motion for spheroidal 
particles with axial symmetry.

\subsection*{Acknowledgements}
M.N.P. acknowledges partial financial support by the ``Supported 
Researcher'' scheme of the University of South Australia and by 
the Max-Planck-Institut f\"ur Metallforschung (MPI-MF) in Stuttgart, 
as well as the hospitality of the MPI-MF Stuttgart. M.N.P. and J.R. 
gratefully acknowledge the financial support from the Australian 
Research Council via the ARC Linkage Grant Scheme and from AMIRA 
International.

\appendix

\section{Derivation of the phoretic velocity}
\label{app_A}

The flow field $\mathbf{u}(\mathbf{r})$ in the outer region is the 
solution of the incompressible, force free Stokes equations
\begin{equation}
\label{St_eq}
\nabla \cdot \mathbf{\hat \Pi} = 0\,,
~\nabla \cdot \mathbf{u} = 0\,,
\end{equation}
subject to the boundary conditions
\begin{equation}
\label{BC_flow}
\left.\mathbf{u}\right|_{\Sigma_\delta}=
\mathbf{V} + \mathbf{v}_s\,,~~
\left.\mathbf{u}\right|_{|\mathbf{r}|\to \infty} = 0\,.
\end{equation}
$\mathbf{\hat \Pi} := -p \mathbf{\hat I} + \mu \mathbf{\hat S}$ is 
the corresponding pressure tensor, where $p$ is the hydrostatic 
pressure and $\mathbf{\hat S}$ is the shear
stress tensor, i.e., $S_{\alpha \beta} =
\partial u_{\alpha}/\partial x_{\beta} +
\partial u_{\beta}/\partial x_{\alpha}$. 
Owing to the linearity of the Stokes equations, we can write the
solution as $\mathbf{u} = \mathbf{u}' + \mathbf{u}''$ and 
$\mathbf{\hat \Pi} = \mathbf{\hat \Pi}' + \mathbf{\hat \Pi}''$,
where $\mathbf{u}'$ and $\mathbf{u}''$ are the solutions of the
incompressible Stokes equations which vanish at infinity and satisfy
the boundary conditions
\begin{equation}
\label{BC_auxiliary}
\left.\mathbf{u}'\right|_{\Sigma_\delta}= \mathbf{v}_s\,
~\mathrm{and}~
\left.\mathbf{u}''\right|_{\Sigma_\delta}=
\mathbf{V}\,,
\end{equation}
respectively, while $\mathbf{\hat \Pi}'$ and $\mathbf{\hat \Pi}''$ 
are the corresponding pressure tensors. By using the condition that 
the motion of the particle and the outer hydrodynamic flow are such 
that there is no net force acting on the system composed of the 
particle plus the surface film, and by replacing 
$\mathbf{\hat \Pi}''$ with the expression for the pressure 
tensor on the surface of an ellipsoid translating through an 
unbounded fluid at rest \cite{Brenner_64},
\begin{equation}
\label{def_tensor_K}
\left(\mathbf{\hat n} \,\mathbf{\hat \Pi}'' \right)_{\Sigma_\delta}= 
- \dfrac{1}{3}\,\dfrac{\mu}{{\cal V}_p} \, 
(\mathbf{\hat n} \cdot \mathbf{r})_{\Sigma_\delta}\, 
\mathbf{\hat K}^{(t)} \,\mathbf{V}\,,
\end{equation}
one obtains
\begin{equation}
\label{force_equality}
\iint\limits_{\Sigma_\delta} d\Sigma \,\mathbf{\hat n} \,
\mathbf{\hat \Pi}' 
= - \iint\limits_{\Sigma_\delta} d\Sigma \,\mathbf{\hat n} \,
\mathbf{\hat \Pi}'' 
= \mu \,\mathbf{\hat K}^{(t)} \,\mathbf{V}\,.
\end{equation}
$\mathbf{\hat n}$ is the unit vector of the direction normal 
to $\Sigma_\delta$ (oriented towards the fluid) and 
$\mathbf{\hat K}^{(t)}$ is a constant diagonal tensor which depends 
on the diameters of the particle only; its explicit expression is 
not needed in the following. 

On the other hand, because the pairs 
$(\mathbf{u}'\,,\mathbf{\hat \Pi}')$ and
$(\mathbf{u}''\,,\mathbf{\hat \Pi}'')$ are, by construction, 
solutions of the force-free, incompressible Stokes equations and 
decay at infinity, they 
satisfy Brenner's (or Lorentz's) reciprocal 
theorem \cite[(b)]{Happel_book}:
\begin{equation}
\label{reciprocal_relation}
\iint\limits_{\Sigma_\delta} d\Sigma \,\mathbf{\hat n} \,
\mathbf{\hat \Pi}' \, \mathbf{u}''
= \iint\limits_{\Sigma_\delta} d\Sigma \,\mathbf{\hat n} \,
\mathbf{\hat \Pi}''\, \mathbf{u}'\,.
\end{equation}
By using Eq. (\ref{def_tensor_K}) to replace $\mathbf{\hat \Pi}''$ 
and the BCs in Eq. (\ref{BC_auxiliary}) to replace $\mathbf{u}'$ 
and $\mathbf{u}''$, and by noting that $\mathbf{V}$ and 
$\mathbf{\hat K}^{(t)}$ are constant with respect to the 
integrations, one obtains
\begin{equation}
\label{V_relations_Brenner}
\mathbf{V}\, 
\iint\limits_{\Sigma_\delta} d\Sigma \,\mathbf{\hat n} \,
\mathbf{\hat \Pi}' = 
- \dfrac{1}{3}\,\dfrac{\mu}{{\cal V}_p} \, 
\mathbf{\hat K}^{(t)} \,\mathbf{V}\,
\iint\limits_{\Sigma_\delta} d\Sigma \,(\mathbf{\hat n} \cdot
\mathbf{r})\,\mathbf{v}_s\,.
\end{equation}
By using Eq. (\ref{force_equality}) for the rhs of 
Eq. (\ref{V_relations_Brenner}) and then replacing in the 
calculations $\Sigma_\delta$ by $\Sigma$, one obtains 
Eq. (\ref{velocity}) for the phoretic velocity.

\section{Phoretic velocity of a spherical Janus particle}
\label{app_B}
The case of a spherical particle ($s_r = 1$) with a catalytic cap
centered at a point chosen as one of its poles (see Fig. \ref{fig1}) 
has been discussed in Ref. \cite{Golestanian_2007}. Thus we provide
here only a brief summary of the results for reasons of completeness, 
further referencing, and provision of some remarks. 
Equation (\ref{diff_eq}) subject to the BCs given in Eq. (\ref{BC}) 
can be conveniently solved using the standard polar spherical 
coordinates $(r,\theta,\phi)$. The solution $\rho(r,\theta)$ of 
Eq. (\ref{diff_eq}) (there is no dependence on $\phi$ because the 
system has azimuthal symmetry) which satisfies the BC in 
Eq. (\ref{farfield_BC}) is
\begin{equation}
\label{series_rho_sphere}
\rho(r,\theta) = \sum_{\ell \geq 0}
\dfrac{d_\ell}{r^{\ell+1}} P_\ell(\cos \theta)\,,
\end{equation}
where the coefficients $d_\ell$ are determined by the BC in
Eq. (\ref{normalJ_BC}). Since the cap-like region on
$\Sigma$ covered by the catalyst can be parameterized by
$(r = R_1, \theta_0 \leq \theta \leq \pi, 0 \leq \phi < 2\pi)$, where
\begin{equation}
\label{def_theta0}
\cos\theta_0 = -1 + s_h\,,
\end{equation}
one obtains
\begin{eqnarray}
\label{expr_d_ell_sphere}
d_\ell
&=&\dfrac{2 \ell +1}{2(\ell+1)}\,\dfrac{R_1^{\ell+2}}{D} \,
\int\limits_{0}^{\pi} d\theta\,(\sin\theta)\,\nu_B\,\sigma\,
\Upsilon(\theta,\theta_0)\,
P_\ell(\cos\theta)\nonumber\\
&=& \dfrac{\nu_B \sigma R_1^{\ell+2}}{D} \,
\dfrac{2 \ell +1}{2(\ell+1)}\,\int\limits_{-1}^{+1} dx\,
\Upsilon(x,\cos\theta_0)\,P_\ell(x) \nonumber\\
&=:&
\dfrac{\nu_B \sigma R_1^{\ell+2}}{D} \,{\tilde d}_\ell(s_h)\,.
\end{eqnarray}
With $\mathbf{\hat n} \cdot \mathbf{r} = r$ and 
$\mathbf{\hat e}_z \cdot \nabla^\Sigma \rho(R_1,\theta)
= - (\sin \theta/R_1) \,\partial_\theta \rho(R_1,\theta)$,
and by using the expressions in Eqs. (\ref{velocity}) and 
(\ref{normalJ_BC}) for the phoretic velocity and for the 
characteristic function $\Upsilon(\mathbf{r})$, respectively, one 
finds:
\begin{eqnarray}
\label{velocity_sphere}
V_{sph} &=& \dfrac{b}{R_1} \,\int\limits_{0}^{\pi} d\theta\, 
\sin \theta \,\cos \theta \, \rho(R_1,\theta)\nonumber\\
&=&
\dfrac{2}{3} V_0 {\tilde d}_1(s_h) = \dfrac{(1-s_h)^2 -1}{4} 
V_0\,.
\end{eqnarray}
In Eq. (\ref{velocity_sphere}) the first equality has been derived in 
Refs. \cite{Golestanian_2005,Golestanian_2007}. The second equality 
follows by using the series expansion in Eq. (\ref{series_rho_sphere}) 
for $\rho(R_1,\theta)$ and due to $\cos \theta = P_1(\cos \theta)$ so 
that in the expansion all terms with $\ell \neq 1$ vanish. We note 
that, as for the general prolate or oblate spheroid, there is no 
explicit dependence on $R_1$ \cite{Golestanian_2007}. 
Equation (\ref{velocity_sphere}) provides the following 
conclusions:\newline
(i) Since $0 \leq s_h \leq 2$, $V_{sph}$ and $V_0$ have opposite signs. 
Thus for repulsive interactions between the particle and the product 
molecules the motion will be in the positive $z$-direction because 
$b$ and thus $V_0$ are negative 
\cite{Anderson_1989,Golestanian_2005}; similarly, for attractive 
interactions, the motion will be in the negative $z$-direction 
(see Fig. \ref{fig1}).\newline
(ii) As expected on basis of symmetry considerations, the maximum 
velocity $V_{sph}^{\mathrm{max}} = - V_0/4$ occurs for $s_h = 1$, 
i.e., if the catalyst covers just a hemisphere.\newline
(iii) For any finite value $\sigma$ of the density of catalytic 
sites the phoretic velocity vanishes if the surface area covered by 
the catalyst tends to zero, i.e., 
$V_{sph}(s_h \to 0; \sigma < \infty) \to 0$, as expected intuitively.
\newline
(iv) Since $2 \pi R_1^2 s_h \sigma$ is the total number of catalytic 
sites in the area covered by the catalyst, the case of a sphere
with a single catalytic source corresponds to the limit
$\{s_h \to 0,~\sigma \to \infty\}$ such that $2 \pi R_1^2 s_h \sigma 
= 1$. The phoretic velocity in this case is
\begin{equation}
V_{sph}(\{s_h \to 0,~\sigma \to \infty, 2 \pi R_1^2 s_h \sigma = 1\})
= - \dfrac{b \nu_B}{4 \pi R^2 D}\,,\nonumber
\end{equation}
in agreement with Ref. \cite{Golestanian_2005}.\newline
(v) If a general axially symmetric distribution of catalytic activities
$\nu_B \mapsto \nu_B f(\theta)$ is considered, as in 
Ref. \cite{Golestanian_2007}, the second equality in 
Eq. (\ref{velocity_sphere}) still holds with the coefficients
${\tilde d}_\ell$ redefined by the particular choice of the 
distribution considered by replacing 
$\Upsilon(\theta) \mapsto f(\theta) \, \Upsilon(\theta)$ in the 
first equation in Eq. (\ref{expr_d_ell_sphere}).

\section{Numerical analysis of the velocity in the limiting case of 
a needle-like particle}
\label{app_C}

According to Eq. (\ref{V_prol_rod_first_approx}), the behavior of 
the velocity of a prolate object in the limit $\xi_0 \to 1^+$ is 
determined by the series
\begin{equation}
\label{f_tilde_series}
{\tilde f}(\xi_0) := \sum_{\ell~\textrm{odd}} (2 \ell +1)
\left[\dfrac{Q_\ell(\xi_0)}{Q'_\ell(\xi_0)} 
\gamma_\ell(\xi_0,\eta_0 = 0)\right]_{\xi_0 \gtrsim 1}\,.
\end{equation} 
Because $Q_\ell(\xi_0 \to 1^+) \sim \ln(\xi_0-1)$ [see 
Eq. (\ref{Ql_def})], one may expect that for $\xi_0 \gtrsim 1$ 
the series above varies as
\begin{eqnarray}
\label{f_tilde_limit}
{\tilde f}(\xi_0 \to 1^+) 
&\simeq& (\xi_0-1) \ln(\xi_0-1)\nonumber\\ 
& \times &\sum_{\ell~\textrm{odd}} 
(2 \ell +1) \gamma_\ell(\xi_0 = 1,\eta_0 = 0) \nonumber\\
&=:& 
(\xi_0-1) \ln(\xi_0-1) f(\xi_0 = 1)\,,
\end{eqnarray}
where the series defining the prefactor $f(1) \equiv f(\xi_0 = 1)$ 
has to be calculated numerically. However, within 
the limits of numerical accuracy it turns out that $f(1) = 0$ so 
that Eq. (\ref{f_tilde_limit}) does not capture the leading behavior 
of ${\tilde f} (\xi_0 \to 1^+)$. Most likely, the reason for the 
failure of this approximation is that the series 
representation of the product density 
[Eq. (\ref{series_rho_prolate})] is not uniformly 
convergent  (and actually completely breaks down) at $\xi_0 = 1$, 
where the differential equation of the Legendre functions is 
singular and $Q_\ell(\xi_0)$ diverges; consequently, there is 
no warranty that the limit $\xi_0 \to 1^+$ can be taken term by 
term. The result $f(1) = 0$ thus suggests that instead the series 
has to be first summed up for general $\xi_0$ and only then the sum 
can be evaluated in the limit $\xi_0 \to 1^+$. 
(See also Ref. \cite{Paxton_2004}, in which the 
thin rod limit of a vanishing ratio between the radius and the 
length of a cylinder [i.e., Eq. (4) therein] could be taken only 
after calculating the velocity as a function of this ratio 
[Eq. (3) therein]; similar arguments apply to the derivation of Eq. (16) 
from (14) in Ref. \cite{Golestanian_2007}.) 

We therefore proceed with a numerical study of the series 
${\tilde f}(\xi_0)$ [Eq. (\ref{f_tilde_series})]
as a function of $\xi_0 \gtrsim 1$. 
By using the recursion relation satisfied by the Legendre 
$Q_\ell$ functions \cite{Hobson_book}
\begin{equation}
\label{Q_ell_recurs}
Q'_\ell(w) = \dfrac{\ell + 1}{w^2-1} [Q_{\ell+1}(w)-w \, Q_\ell(w)]
\end{equation}
and the series representation [Eq. (\ref{Ql_def})] for $Q_\ell(w)$, 
one concludes that for $\xi_0 > 1$ and $\ell \gg 1$ the behavior 
of the ratio involving the Legendre functions $Q_\ell$ is given by
\begin{equation}
\label{Q_ell_ratio_asymp}
\left[\dfrac{Q_\ell(\xi_0)}{Q'_\ell(\xi_0)} 
\right]_{\ell \gg 1} 
\sim \dfrac{2 \xi_0 \, (\xi_0^2 - 1)}{(1 - 2 \xi_0^2)} 
\,\dfrac{1}{\ell+1} < 0\,.
\end{equation}
Turning now to the behavior of $I_\ell(\xi_0):= 
\gamma_\ell(\xi_0, \eta_0 = 0)$ for $\xi_0 \gtrsim 1$, we first 
note that [Eq. (\ref{gamma_ell})]
\begin{eqnarray}
\label{bound_gamma}
|I_\ell(\xi_0)|&\leq& \int\limits_{-1}^{0} \, d\eta\, 
\left|\sqrt{\xi_0^2 - \eta^2} P_\ell(\eta)\right| \nonumber\\
&\leq& \xi_0 \int\limits_{-1}^{0} \, d\eta\, |P_\ell(\eta)| 
\leq \xi_0\,.
\end{eqnarray}
For large $\ell$, i.e., $\ell > 50$ and values of $\xi_0$ very close 
to 1, i.e., $\xi_0 -1 \leq 10^{-6}$, an accurate direct numerical 
determination of $I_\ell(\xi_0)$ [Eq. (\ref{gamma_ell})] is extremely 
difficult mainly because of the oscillatory behavior of the Legendre 
polynomials, and one thus has to find a way to reformulate the 
integral. To this end we make use of the representation of the 
Legendre polynomials with \textit{odd} index $\ell$ in terms of the 
hypergeometric function $_2F_1$ \cite{Hobson_book}:
\begin{eqnarray}
\label{LegendreP_to_hypergeometric}
&&P_\ell(\eta) = (-1)^{\frac{\ell-1}{2}} 
\dfrac{\ell!}{2^{\ell-1} \left(\dfrac{\ell-1}{2} !\right)^2} 
\nonumber\\
&& \times \eta\,\,_2F_1
\left(-\dfrac{\ell-1}{2},\dfrac{\ell}{2}+1;\dfrac{3}{2};\eta^2\right)\, 
:=  \alpha_\ell \,\,p_\ell(\eta)\,, ~~
\end{eqnarray}
where $\alpha_\ell$ denotes the prefactor and 
\newline
$p_\ell(\eta) = \eta\,
_2F_1
\left(-\dfrac{\ell-1}{2},\dfrac{\ell}{2}+1;\dfrac{3}{2};\eta^2\right)$ 
is a polynomial of order $\ell$. The hypergeometric 
function $_2F_1(w)$ obeys the relation \cite[(d)]{Abramowitz_book}
\begin{eqnarray}
\label{derivative_hypergeometric}
&&\dfrac{d}{d w}\, _2F_1 \left(a,b;c;w^2\right) = 
2 \,\dfrac{a \,b}{c} \,w\, \nonumber\\ 
&&~~~~~~~~~~~~~\times _2F_1 \left(a+1,b+1;c+1;w^2\right)\,,
\end{eqnarray}
with $_2F_1 \left(0,b;c \neq 0;w^2\right) \equiv 1$ and
$ _2F_1 (a,b;c;0) = 1$, and it is well defined, as well as its 
derivatives, at $w = 1$ (being polynomials). Accordingly, for $\ell$ 
odd,  $I_{\ell}(\xi_0)$ can be computed via successive integrations 
by parts:
\begin{eqnarray}
\label{integr_by_parts}
I_{\ell}(\xi_0)/\alpha_\ell 
&=& 
\int\limits_{-1}^{0} d\eta \, 
\left[ -\dfrac{1}{3} (\xi_0^2-\eta^2)^{3/2}\right]' \nonumber\\
&\times&{_2F_1}
\left(-\dfrac{\ell-1}{2},\dfrac{\ell}{2}+1;\dfrac{3}{2};\eta^2\right) 
\nonumber\\
&=& -\dfrac{1}{3} \xi_0^3 + a_1(\ell,0) (\xi_0^2-1)^{3/2} \nonumber\\
&+&\left(\dfrac{2}{3}\right)^2 
\left(\dfrac{1}{2}-\dfrac{\ell}{2}\right)\left(1+\dfrac{\ell}{2}\right)
\nonumber\\
&\times&
\int\limits_{-1}^{0} d\eta \left[ -\dfrac{1}{5} 
(\xi_0^2-\eta^2)^{5/2}\right]' \nonumber\\
&\times&{_2F_1}
\left(
-\dfrac{\ell-1}{2}+1,\dfrac{\ell}{2}+1+1;\dfrac{3}{2}+1;\eta^2
\right)
\nonumber\\
&=&-\dfrac{1}{3} \xi_0^3-\left(\dfrac{2}{3}\right)^2 
\dfrac{1}{5}\left(\dfrac{1}{2}-\dfrac{\ell}{2}\right)
\left(1+\dfrac{\ell}{2}\right)\xi_0^5
\nonumber\\ 
&+& a_1(\ell,0) (\xi_0^2-1)^{3/2} +  
a_1(\ell,1) (\xi_0^2-1)^{5/2}\nonumber\\ 
&+& 
\left(\dfrac{2}{3}\right)^2 \left(\dfrac{2}{5}\right)^2 
\left(\dfrac{1}{2}-\dfrac{\ell}{2}\right) 
\left(\dfrac{3}{2}-\dfrac{\ell}{2}\right)\nonumber\\
&\times&
\left(1+\dfrac{\ell}{2}\right)\left(2+\dfrac{\ell}{2}\right)
\nonumber\\
&\times&
\int\limits_{-1}^{0} d\eta 
\left[ -\dfrac{1}{7} (\xi_0^2-\eta^2)^{7/2}\right]' \nonumber\\
&\times&{_2F_1}
\left(
-\dfrac{\ell-1}{2}+2,\dfrac{\ell}{2}+1+2;\dfrac{3}{2}+2;\eta^2
\right)
\nonumber\\
& = & \cdots \mathrm{~in~total~}(\ell-1)/2 
\mathrm{~steps\,,}\nonumber\\
&&\mathrm{~until~reaching}~{_2F_1}(0,b;c;\eta^2) \equiv 1 \,.
\end{eqnarray}
The coefficients $a_1(\ell,k)$ of the terms 
$(\xi_0^2-\eta^2)^{3/2+k}$ are proportional to 
${_2F_1}
\left(
-\dfrac{\ell-1}{2}+k,\dfrac{\ell}{2}+1+k;\dfrac{3}{2}+k;1
\right)$ and can be determined analytically. 
The terms formed by powers of $\xi_0$ can be summed up in closed form, 
and it turns out that they provide a very good approximation for the 
value of $I_\ell(\xi_0)$:
\begin{eqnarray}
\label{closed_sum_powers_xi0}
I_{\ell}(\xi_0 &\gtrsim& 1) \simeq \alpha_\ell
\left\lbrace-\dfrac{1}{3} \xi_0^3 -
\sum_{s = 0}^{(\ell-3)/2} 
\left[\prod_{m = 0}^{s}\left(\dfrac{2}{2 m +3}\right)^2\right] \right.
\nonumber\\ 
&\times& \left.\left[\prod_{m = 0}^{s}\left(\dfrac{1}{2}- 
\dfrac{\ell}{2} +m \right) 
\left(1+\dfrac{\ell}{2} +m \right)\right]\,
\dfrac{\xi_0^{2 s + 5}}{2 s + 5}\right\rbrace\nonumber\\
&\simeq& 
\alpha_\ell
\left[-\dfrac{1}{3} \xi_0^3 + \dfrac{\ell^2+\ell-2}{45}\, \xi_0^{5} 
\right.\nonumber\\
&\times& \left._{3}F_{2} 
\left(\left\lbrace 1,\dfrac{3}{2}-\dfrac{\ell}{2},2+\dfrac{\ell}{2} 
\right\rbrace;
\left\lbrace\dfrac{5}{2},\dfrac{7}{2}\right\rbrace;\xi_0^2\right)
\right]\,.
\end{eqnarray}
Since Eq. (\ref{closed_sum_powers_xi0}), which leads to a polynomial of 
order $2 \dfrac{\ell-3}{2} + 5 = \ell+2$ [see the 
first equality in Eq. (\ref{closed_sum_powers_xi0})], predicts that 
at large $\ell$ and $\xi_0 > 1$ the behavior is dominated by the 
term $\xi_0^{\ell+2}$, it obviously represents an approximation 
which breaks down for very large $\ell$ or for not small enough 
values of $\xi_0-1$. However, this breakdown of the approximation 
in Eq. (\ref{closed_sum_powers_xi0}) can be easily monitored by 
checking that the result obeys the bound given by Eq. 
(\ref{bound_gamma}). The approximation can be systematically 
improved, if needed, by including the terms $a_1(\ell,k) 
(\xi_0^2-\eta^2)^{3/2+k}$ (which can be computed analytically) as 
they become relevant. For $\xi_0 - 1 \leq 10^{-4}$ and 
$\ell \leq 201$ ($\ell$ odd), we did not have to add any such 
corrections to Eq. (\ref{closed_sum_powers_xi0}).

The closed form of Eq. (\ref{closed_sum_powers_xi0}) significantly 
simplifies the numerical study of the behavior of 
$I_{\ell}(\xi_0 \gtrsim 1)$ (with $\ell$ odd). It strongly supports 
(by extrapolating the results within the range where it provides an 
accurate approximation) that $I_{\ell \to \infty}(\xi_0) \to 0$ and 
that $\mathrm{sign}\left[I_{\ell}(\xi_0)\right] = 
\mathrm{sign} (\alpha_\ell) = (-1)^{\frac{\ell-1}{2}}$. In 
combination with Eq. (\ref{Q_ell_ratio_asymp}), this shows that 
approximating $I_{\ell}(\xi_0 \gtrsim 1)$ by Eq. 
(\ref{closed_sum_powers_xi0}) in the series representation of 
Eq. (\ref{f_tilde_series}) leads to a general form of the term in 
the series for ${\tilde f}(\xi_0)$ which has alternating sign and 
vanishes with increasing $\ell$, which ensures that as expected 
the series is convergent.

We define the partial sum ${\tilde f}_L(\xi_0)$, with $L$ odd, 
of the series in Eq. (\ref{f_tilde_series}) as
\begin{eqnarray}
\label{part_sum}
{\tilde f}_L(\xi_0) &=& 
\dfrac{1}{2} 
\left(
\sum_{\ell = 1,~\ell~\textrm{odd}}^{L} (2 \ell +1)
\dfrac{Q_\ell(\xi_0)}{Q'_\ell(\xi_0)} 
I_\ell(\xi_0)\right.\nonumber\\
&+& \left. \sum_{\ell = 1,~\ell~\textrm{odd}}^{L + 2} (2 \ell +1)
\dfrac{Q_\ell(\xi_0)}{Q'_\ell(\xi_0)} 
I_\ell(\xi_0)
\right)\,,
\end{eqnarray}
such that the oscillatory behavior induced by the alternating signs 
of successive terms is damped. Guided by the logarithmic divergence 
of $Q_\ell(\xi_0 \to 1^+)$, by the fact that the series defining 
the prefactor $f(1)$ [Eqs. (\ref{f_tilde_series}) and 
(\ref{f_tilde_limit})] turned out to be convergent (although 
vanishing), and by the results for the phoretic velocity of a 
rod-like particle in Refs. 
\cite{Golestanian_2007,Paxton_2004}, we make the \textit{ansatz} 
that ${\tilde f}(\xi_0)$ behaves as 
 \begin{equation}
\label{f_tilde_ansatz}
{\tilde f}(\xi_0 \gtrsim 1) = \underbrace
{(\xi_0-1) \ln(\xi_0-1)}_{~~~~~~~~:=\,\,r(\xi_0)} f(\xi_0)
\end{equation}
and aim at determining numerically the function $f(\xi_0)$.

\begin{figure}[!htb]
\begin{center}
\includegraphics[width=.9 \linewidth]{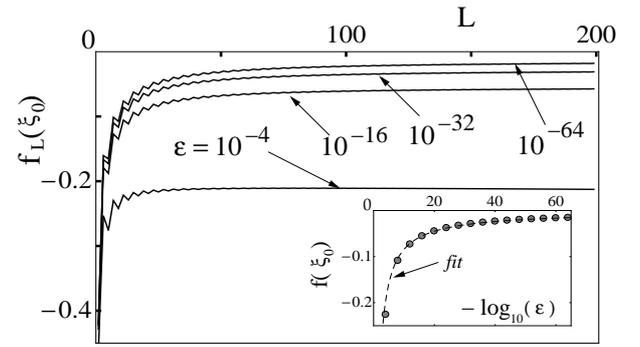}
\end{center}
\caption
{
\label{fig5}
Dependence of $f_L(\xi_0)$ (see the main text) on $L$ (for $L$ odd) 
for $\epsilon:= \xi_0 - 1 = 10^{-4}$, $10^{-16}$, $10^{-32}$, 
and $10^{-64}$. The inset shows the resulting $f(\xi_0)$ (points) 
obtained from the extrapolated $f_{L \to \infty} (\xi_0)$, as well 
as the power law fit $-0.7 \times [-\log_{10}(\epsilon)]^{-0.9}$ 
(dashed line), as a function of $- \log_{10}(\epsilon)$. 
}
\end{figure}

In Fig. \ref{fig5} we show the dependence of \newline
$f_L(\xi_0) = {\tilde f}_L(\xi_0)/r(\xi_0)$ on $L$ for various values 
$\epsilon := \xi_0 -1$, $10^{-64} \leq \epsilon \leq 10^{-4}$. The 
convergence of the partial sums is clear, and this allows us to 
extract the values $f(\xi_0)$ as the corresponding constant plateau 
values of each of the curves  $f_L(\xi_0)$. The numerically 
determined $f(\xi_0)$, for clarity shown in the inset of 
Fig. \ref{fig5} as a function of $-\log_{10}(\epsilon)$, highlights
the following important features: \textbf(i) $f(\xi_0) < 0$ for all 
$\xi_0 \gtrsim 1$, and \textbf(ii) the data set $\lbrace 
-\log_{10}(\epsilon),f(\xi_0 = 1+\epsilon)\rbrace$ is very well 
fitted by a power law [dashed line in the inset in Fig. \ref{fig5}] 
leading to $f(\xi_0 \to 1^+) \simeq -0.7\,\,[-\log_{10}(\xi_0-1)]^{-0.9} 
\simeq -1.5\,\,[-\ln(\xi_0-1)]^{-0.9}$.
%%%%%%%%%%%%%%%%%%%%%%%%%
%%%%%%%%%%%%%%%%%%%%%%%%%



\begin{thebibliography}{0}
\bibitem{note_2}
Besides the translation described by $\mathbf{V}$, in the most 
general case a term accounting for a rigid-body rotation of the 
particle with angular velocity $\mathbf{\Omega}$ should also be 
considered. However, this angular velocity turns out to be 
identically zero in most cases in which the particle has 
homogeneous surface properties \cite{Anderson_1989,Morrison_1970}.
The azimuthal symmetry of our system and the additional assumption
that the properties of the catalyst-covered surface are similar to
those of the inert part (as far as the particle-solute effective 
interaction is concerned) ensures that we are dealing with such a 
case; therefore we disregard the rotation term here.
\bibitem{note_3} 
The numerical calculations have been performed using the software 
Mathematica (version 7.01), for which the Legendre function 
$Q_\ell(x)$ for arguments $x > 1$ or 
$x \in \mathbb{C} \setminus \mathbb{R}$ is implemented as ``the 
Legendre function Q of type 3'' \textsf{LegendreQ[$\ell$,0,$x$,3]}.
\end{thebibliography}
\end{document}